\documentclass[a4paper,12pt]{article}
\usepackage[latin2]{inputenc}
\pdfoutput=1
\usepackage{t1enc}
\usepackage{anysize}
\usepackage{amsmath}
\usepackage{amsfonts}
\usepackage{amssymb}
\usepackage{graphicx}
\usepackage{subfigure}
\usepackage{multirow}
\usepackage{booktabs}
\usepackage{listings}
\usepackage{float}
\usepackage{listings}
\usepackage{bm}
\usepackage{url}
\usepackage{authblk}
\usepackage[table,xcdraw]{xcolor}
\usepackage{physics}
\usepackage{fancyvrb}
\usepackage{enumitem}

\newlist{step}{enumerate}{1}
\setlist[step]{label=Step \arabic*:}

\marginsize{2.5cm}{2.5cm}{2.5cm}{2.5cm}
\usepackage{ntheorem}
\theoremstyle{change}
\makeatletter
\theoremseparator{.}
\newtheoremstyle{magyartetelstilus}%
{\item[\hskip\labelsep\theorem@headerfont
            ##2.\ ##1\theorem@separator]}%
{\item[\hskip\labelsep\theorem@headerfont
            ##2.\ ##1\theorem@separator\ (##3)]}%
\theoremstyle{magyartetelstilus}
\newtheorem{tetel}{Theorem}[section]

\newtheorem{áll}[tetel]{Proposition}
{\theorembodyfont{\rmfamily}

}%

\DeclareMathOperator{\tg}{\mathrm{tan}}
\DeclareMathOperator{\sgn}{sign}

\renewcommand{\vec}[1]{\mathbf{#1}}
\makeatother

\begin{document}
\thispagestyle{empty}

\title{Multivariate stable distributions and their applications for modelling cryptocurrency-returns}

\author{Szabolcs Majoros\textsuperscript{a} and András Zempléni\textsuperscript{b}\thanks{CONTACT zempleni@caesar.elte.hu}
\\ \small{{\textsuperscript{a,b}E\"otv\"os Lor\'and University,  Institute of Mathematics, Department of Probability Theory and Statistics, Budapest, Hungary}}}
\date{}
\maketitle

\noindent\makebox[\linewidth]{\rule{\paperwidth}{0.4pt}}

\begin{abstract}
In this paper we extend the known methodology for fitting stable distributions to the multivariate case and apply the suggested method to the modelling of daily cryptocurrency-return data. The investigated time period is cut into 10 non-overlapping sections, thus the changes can also be observed. We apply bootstrap tests for checking the models and compare our approach to the more traditional extreme-value and copula models.
\\[1mm]
\end{abstract}

\section{Introduction}\label{intr}

Modelling the price fluctuations of the cryptocurrencies, which behave rather erratically, providing the chance for huge gains within a short period -- together with the possibility of similar losses is a major challenge and definitely of interest for the potential investors as well as for the theoreticians. One may find some preliminary, mostly descriptive statistics-based calculations, like \cite{19}. The available amount of data is not huge, but it may just be enough for some preliminary two- or three-dimensional modelling. 

There are some standard methods for univariate modelling of heavy tailed distributions: the extreme-value distributions may be used for separate models for the gains and the losses either by block-maxima or peaks-over-threshold models (see e.g. Pickands, \cite{Pi}, or the summary work of Embrechts et al. \cite{14} and the references therein).  However, a joint model for both tails would definitely to be preferred. The same applies for the multivariate approaches, see Coles and Tawn \cite{co} for an early work. A standard copula model -- by e.g. one of the models in Nelsen \cite{ne} -- looks as appealing, but again, these parametric families may not be suitable for capturing the unusual dependencies among the cryptocurrencies. Nonparametric copula estimators are not much simpler than the methods we propose, but without the theoretical background of the multivariate stable distributions.
  
Stable distributions are a rich class of probability distributions. Paul Lévy was the one who first studied this distribution family and he proved the Generalized Central Limit Theorem, which gives the theoretical background for their use in modelling (\cite{le}). Their application goes back to Benoit Mandelbrot, who was modelling cotton price changes with stable distributions \cite{16}. Since then many other studies were published in the subject. The theory was further developed in many papers of Zolotarev (see e.g. \cite{zol}). The most recent work was written by J. P. Nolan \cite{1}. 

The paper is structured as follows:
At first we  define multivariate stable distributions and show their most important properties in Section 1.1. 

In Section 2 we introduce their parameter estimation methods, first in the univariate, then in the bivariate case. The problem is rather challenging, as their density function does not have a closed form. 
We also propose a new general estimation method, applicable to higher dimensions. 

In Section 3, we apply our methods to the exciting new financial instruments, the daily logreturns of the three most important cryptocurrencies. In Section 4 we compare the results to more traditional modelling tools like generalized Pareto distributions or copulas and conclude the paper by a short Summary.

\subsection{Stable distributions}\label{int_uni}

The following introduction to stable distributions is based on \cite{1}.
By definition, a $d$-variate random variable $X$ is stable, if to any positive $a,b\in\mathbb{R}$, there are positive $c$ and $d\in\mathbb{R}^d$, such that  $F_{aX_1+bX_2}=F_{cX+d}$, where the random variables $X_1$, $X_2$ and $X$ are i.i.d. We may be familiar with this property from the normal distribution, as it is a member of the stable distribution family too.

In the univariate case, the distribution is described by four parameters: index $\alpha\in(0,2]$, skewness $\beta\in[-1,1]$, scale $\gamma>0$ and shift $\delta\in\mathbb{R}$. The usual notion for the distribution is $S(\alpha,\beta,\gamma,\delta)$. In general there is no closed form of their density function, apart from a few special cases: the well known normal distribution, the Cauchy and Lévy distributions. They are described by their characteristic functions, as follows:
\begin{align*}
\varphi(t)=
\begin{cases}
\exp\left\{-\gamma^\alpha\vert t \vert^\alpha(1+i\beta\tg{\frac{\pi\alpha}{2}}\cdot\sgn{t})(\vert \gamma t \vert^{1-\alpha}-1)+i\delta t\right\} &\quad{\alpha\neq1}\\
\exp\big\{-\gamma\vert t \vert\left(1+i\beta\frac{2}{\pi}\sgn{t}\cdot\log{(\gamma\vert t \vert)}\right)+i\delta t\big\} &\quad{\alpha=1}.
\end{cases}
\end{align*}
Stable distributions are always absolutely continuous and unimodal.

There are a few different parametrisations of the stable distributions. The above form is called the $S_0$ representation, which is the one mostly used in statistical modelling, due to the easy interpretation of the parameters.

We may identify the well-known special cases: $S(2,0,\gamma,\delta)$ gives the normal distribution  $N(\delta,2\gamma^2)$.  $S(1,0,\gamma,\delta)$ is the Cauchy distribution and for $\alpha=0.5$ and $\beta=1$ we get the Lévy distribution.
One of the most interesting properties of stable distributions is that for $\alpha<2$ only the $l<\alpha$ moments are finite. If $\beta=0$, the distribution is symmetric with respect to $\delta$. Also, if $\alpha$ is close to 2, then $\beta$ doesn't have much impact on the skewness. This can be easily seen from the characteristic function, where if we substitute $\alpha=2$, the value of the $\beta\tg{\frac{\pi\alpha}{2}}$ in the characteristic function will be 0, therefore $\beta$ doesn't play any role. 

\begin{figure}[H]
	\centering
		\includegraphics[scale=0.35]{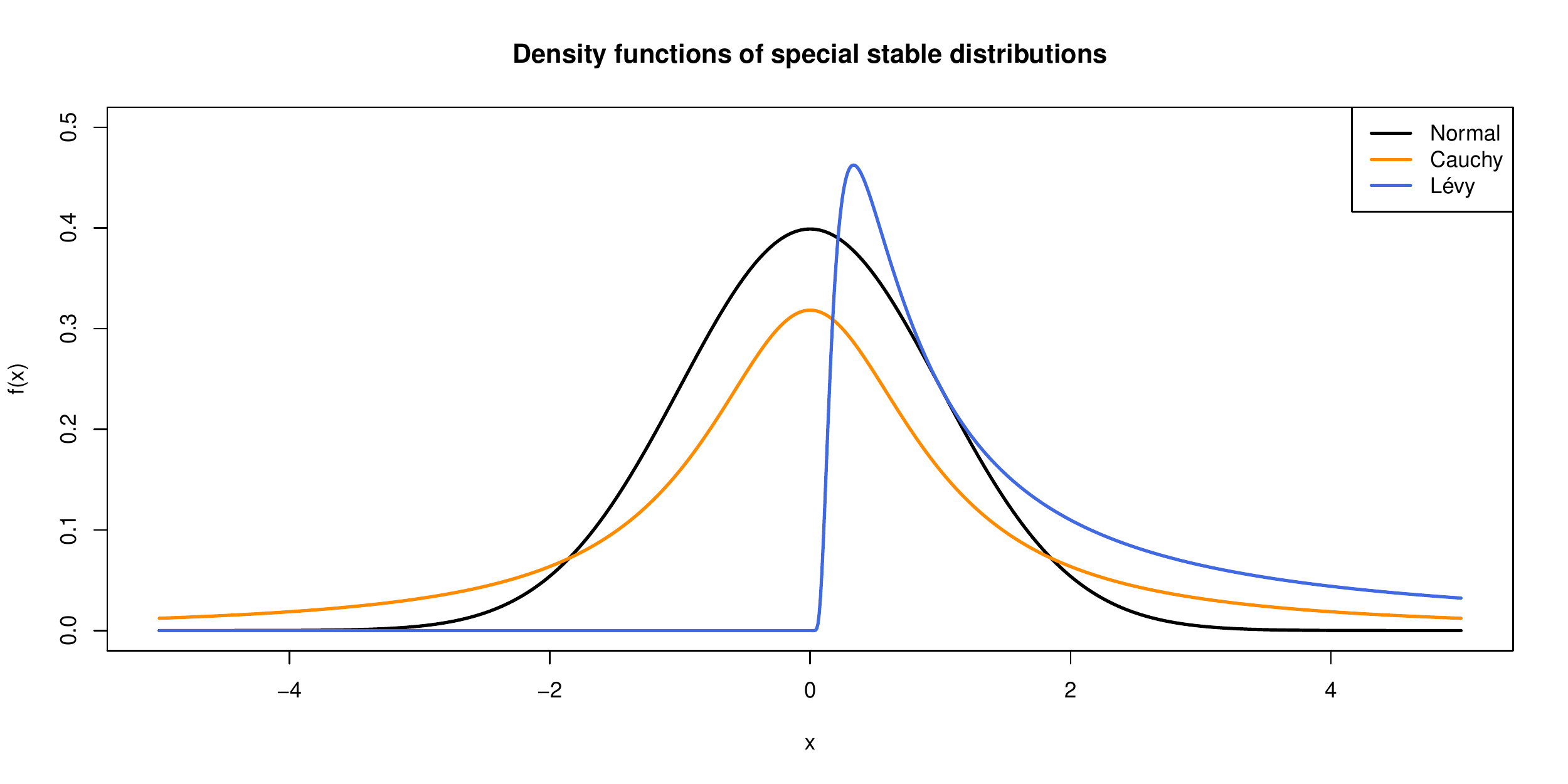}
		\caption{Special stable distributions}
\end{figure}

We can standardize the distribution the same way as we are used to at normal distributions. By dividing a stable $S(\alpha,\beta,\gamma,\delta)$ r.v. by $\gamma$ and subtracting  $\delta$, the distribution will be $S(\alpha,\beta,1,0)$, which may be denoted by $S(\alpha,\beta)$. This makes the distribution family very flexible in practical use.

Now let us turn to the multivariate case. Here the family of stable distributions is nonparametric, as the following characterisation shows.
\begin{tetel}
	Let $\Lambda$ be a finite measure on $S_d$, where $S_d=\left\{ s\in\mathbb{R}^d\colon \Vert s\Vert_2=1\right\}$, the surface of the unit ball. This measure is called the spectral measure. The $d$-dimensional variable $\vec{X}$ is stable, denoted by $\vec{X}\sim S(\alpha,\Lambda,\bm{\delta})$, where $0<\alpha\leq 2$ and $\bm{\delta}\in\mathbb{R}^d$, if and only if its characteristic function is 
	\begin{equation*}
		\varphi_{\vec{X}}(\vec{t})=\exp\{-I_{\vec{X}}(\vec{t})+i\vec{t}^{\mathsf{T}}\bm{\delta}\},\label{mkar1}
	\end{equation*}
	where
	\begin{align*}
		I_{\vec{X}}(\vec{t})=\int_{S_d}\psi\left(\vec{t}^{\mathsf{T}}\vec{s};\alpha\right)\Lambda(d\vec{s})
	\end{align*}
	and
	\begin{align*}
		\psi\left(u;\alpha\right)=
		\begin{cases}
			\lvert u \rvert^{\alpha} \left(1-i\tg\frac{\pi\alpha}{2}\cdot\sgn u\right) &\quad\alpha\neq 1\\
			\lvert u \rvert(1+i\frac{2}{\pi}\sgn u\cdot\log\rvert u \lvert) &\quad\alpha=1.
		\end{cases}
	\end{align*}
	The function $I_{\vec{X}}(\vec{t})$ determines the shape of the distribution and $\bm{\delta}$ is the location vector.
\end{tetel}
As we can see $\alpha$ and $\delta$ essentially remained the same as in the univariate case, which is not true for $\beta$ and $\gamma$. Instead, the measure $\Lambda$ takes over their role. Additionally, this measure is what determines the dependence structure of the distribution, which makes the model fitting more complicated, since its non-parametric estimation is not feasible. Because of this, we propose a parametric model later on, where this $\Lambda$ is discrete, more exactly that $\Lambda$ is concentrated to a finite number of points. In this case, the measure can be written as
\begin{equation*}
	\Lambda(\cdot)=\sum_{i=1}^{n} \lambda_i\delta_{s_i}(\cdot)
\end{equation*}
where $\lambda_i$ are the weights concentrated on $\delta_{\vec{s_i}}$ points of mass, $\vec{s_i}\in S_d$. With the discrete $\Lambda$, the characteristic function of $\vec{X}$ simplifies into the following form:
\begin{align}
	\varphi^*(\vec{t})=\exp \left\{-\sum_{i=1}^{n} \psi(\vec{t}^{\mathsf{T}}\vec{s_i};\alpha)\lambda_i+i\vec{t}^{\mathsf{T}}\bm{\delta} \right\}.\label{mkar2}
\end{align}

There is another important property of this distribution family, namely the stability of the linear combinations of its coordinates.
\begin{áll}\label{proj_tul}
	If $\vec{X}$ is $d$-dimensional stable with $0<\alpha\leq 2$, then for every $u\in\mathbb{R}^d$
	\[
	\vec{u}^{\mathsf{T}}\vec{X}=u_1X_1+\ldots+u_dX_d
	\]
	is a univariate stable random variable, with the same $\alpha$.
\end{áll}
We note that the univariate variables $\vec{u}^{\mathsf{T}}\vec{X}\sim S(\alpha,\beta(\vec{u}),\gamma(\vec{u}),\delta(\vec{u}))$, 
completely determine $\vec{X}$, as it can be seen from the next theorem.
\begin{tetel}
	Let be $\vec{u}^{\mathsf{T}}\vec{X}\sim S(\alpha,\beta(\vec{u}),\gamma(\vec{u}),\delta(\vec{u}))$. Then the parameter functions determining $\vec{X}$ can be written in the following form: 
	\begin{align}
		&\gamma(\vec{u})=\left(\int_{S_d} \lvert\vec{u}^{\mathsf{T}}\vec{s}\rvert^{\alpha}\Lambda(d\vec{s})\right)^{1/\alpha}\label{proj1}\\
		&\beta(\vec{u})=\gamma(\vec{u})^{-\alpha}\int_{S_d} \lvert\vec{u}^{\mathsf{T}}\vec{s}\rvert^{\alpha}\sgn(\vec{u}^{\mathsf{T}}\vec{s})\Lambda(d\vec{s)}\label{proj2}\\
		&\delta(\vec{u})=
		\begin{cases}
			\vec{u}^{\mathsf{T}}\bm{\delta} \quad &\alpha\neq 1\\
			\vec{u}^{\mathsf{T}}\bm{\delta}-\frac{2}{\pi}\int_{S_d} \vec{u}^{\mathsf{T}}\vec{s}\cdot\log(\lvert\vec{u}^{\mathsf{T}}\vec{s}\rvert)\Lambda(d\vec{s}) \quad &\alpha=1\label{proj3}.
		\end{cases}
	\end{align}
	Using these, $I_{\vec{X}}(\vec{t})$ can be written as
	\begin{align}\label{ifgv}
		I_{\vec{X}}(\vec{t})=
		\begin{cases}
			\gamma^{\alpha}(\vec{t})(1-i\beta(\vec{t})\tg\frac{\pi\alpha}{2}) \quad&\alpha\neq 1\\
			\gamma(\vec{t})(1-i\delta(\vec{t})) \quad&\alpha=1.
		\end{cases}
	\end{align}
\end{tetel}
The connection between these properties gives us the opportunity to determine the multivariate distribution using the univariate projections and to perform calculations more easily. These are giving the base of the estimation procedure, which we can see in Section \ref{estim}.

There are some special multivariate stable distributions worth mentioning, even if they will not be present explicitly in the estimation procedure. 

\begin{áll}
	If the components of $\vec{X}=(X_1,X_2,\ldots,X_n)$, $X_i\sim S(\alpha,\beta_i,\gamma_i,\delta_i)$ are independent, then the characteristic function of $\vec{X}$ can be written as
	\begin{equation*}
		\varphi_{\vec{X}}(\vec{t})=\exp \left\{-\sum_{i=1}^{n} \omega(t_i;\alpha,\beta_i)\gamma_i^{\alpha}+i\vec{t}^{\mathsf{T}}\bm{\delta}\right\},
	\end{equation*}
\end{áll}
Thus the case of independent components can be represented by a discrete $\Lambda$, where only the intersection of the hypersphere $S_d$ and the axes have positive weights.

The most important theoretical property of stable distributions is the generalized central limit theorem \cite{1}.
\begin{tetel}\label{genlim}
The  random variable $X$ is stable, where $0<\alpha\leq 2$ if and only if there are non-degenerate, independent, identically distributed random variables $X_1,X_2,\ldots,X_n$  and $a_n,b_n\in\mathbb{R}$ normalizing sequences, so that
\begin{equation*}
\dfrac{X_1+X_2+\ldots+X_n}{b_n}-a_n\overset{d}{\to} X.
\end{equation*}
\end{tetel}
The main difference between the classical central limit theorem and the theorem above is that it doesn't require $X$ to have finite second moment. The consequence of the theorem is that the domain of attraction of stable distributions is not empty. An example for suitable $X_i$ is, if its tails satisfy $x^{\alpha}P(X>x)\to c^{+}$ and $x^{\alpha}P(X<-x)\to c^{-}$ as $x \to \infty$, with $c^{+}+c^{-}>0$ except if $\alpha<1$ and $|\beta|=1$ (see Nolan, \cite{1}).
An analogous theorem is true for the multivariate case as well, here the limit is necessarily a multivariate stable distribution.

\section{Parameter estimation}\label{estim}
Since the density function of stable distributions cannot be given in a closed form, the parameter estimation is a difficult task. Maximum likelihood method can be used for the univariate case, but it is very slow in practice, as it is based on inverting the characteristic function, so we use it just for estimating the most important parameter $\alpha$. The method of moments estimation can't be used, because the moments may not exist. So in this case the quantile method, proposed by McCulloch in \cite{6} is the most common choice. This estimation procedure is based on the sample quantiles, which we can compute easily. Its another advantage is that the ML-estimator for $\alpha$ can easily be incorporated into the equations determining the estimators for the other parameters. 

A bivariate estimation method was proposed in \cite{5}, \cite{2} and \cite{10}, which builds on the distribution's properties mentioned in Section \ref{intr}. The method is based on the univariate projections and characteristic function of the distribution, where it is assumed that $\Lambda$ is discrete. Using these, we get an equation system, whose solution will be the estimation of $\Lambda$.
\subsection{Estimating procedure in two dimensions}
This subsection is based on the work of Nolan et al. \cite{2}.
Let $\vec{X_1},\ldots,\vec{X_m}$ be our bivariate sample, from a bivariate stable distribution. Additionally, we assume that $\Lambda$ is discrete and concentrated exactly on $n$ points. 

\subsubsection*{Step 1}
Firstly, the shift $\bm{\delta}=\begin{bmatrix} \delta_1 \\ \delta_2 \end{bmatrix}$ is eliminated. This correction makes the calculations easier. It can be carried out by estimating $\delta_1$ and $\delta_2$ separately e.g. using the quantile method for both marginal distributions and then subtract $\hat{\delta}_1$ and $\hat{\delta}_2$ from the corresponding margin. We can do this, as shifts do not change the other parameters as we could see in Section \ref{int_uni}. Now the characteristic function looks like as
\[
\varphi_0(\vec{t})=\exp \left\{-\sum_{i=1}^{n} \psi(\vec{t}^{\mathsf{T}}\vec{s_i};\alpha)\lambda_i \right\}.
\]

\subsubsection*{Step 2}
In the next step the points $\vec{s_j}=\left(\cos\left(\frac{2\pi(j-1)}{n}\right),\sin\left(\frac{2\pi(j-1)}{n}\right)\right)$, $j=1,\ldots,n$ are chosen as the support of the spectral density $\Lambda$. These form an equidistant partition of points on the unit circle. Additionally, we need a grid for the characteristic function $\vec{t_1},\ldots,\vec{t_n}\in S_2$, which determines the projections $\langle\vec{t_j},\vec{X_1}\rangle,\ldots,\langle\vec{t_j},\vec{X_m}\rangle$. To make the calculation easier, we take these grid points being identical to $s_j$: $\vec{t_j}=\vec{s_j}$, $j=1,\ldots,n$.

\subsubsection*{Step 3}
For every projection, the value of \eqref{proj1} and \eqref{proj2} is estimated. To be able to do that, we need to use quantile or ML methods to estimate $\alpha$ on the constructed projections. Since $\alpha$ has to be constant, we take a pooled version of the estimator as $\hat{\alpha}^*=\frac{1}{n}\sum_{j=1}^n\hat{\alpha}(\vec{t_j})$. After evaluating \eqref{proj1} and \eqref{proj2} 
we can calculate the estimated values of $I_{\vec{X}}(\vec{t})$ for every projection.

\subsubsection*{Step 4}
Since $\Lambda$ is discrete, $I_{\vec{X}}(\vec{t})$ can be written into the form: $I_{\vec{X}}(\vec{t})=\sum_{j=1}^n\psi\left(\vec{t}^{\mathsf{T}}\vec{s_j};\hat{\alpha}^*\right)\lambda_j$. Based on that, define the $n\times n$ complex matrix $\bm{\Psi}$ as 
\begin{align}
\bm{\Psi}(\vec{t_1},\ldots,\vec{t_n};\vec{s_1},\ldots,\vec{s_n})=
\begin{bmatrix}
\psi\left(\vec{t_1}^{\mathsf{T}}\vec{s_1};\hat{\alpha}^*\right) & \dots & \psi\left(\vec{t_1}^{\mathsf{T}}\vec{s_n};\hat{\alpha}^*\right) \\
\vdots & \ddots & \vdots \\
\psi\left(\vec{t_n}^{\mathsf{T}}\vec{s_1};\hat{\alpha}^*\right) & \dots & \psi\left(\vec{t_n}^{\mathsf{T}}\vec{s_n};\hat{\alpha}^*\right)\label{est_mtx}
\end{bmatrix}
\end{align}
and the $n\times 1$ unknown vector $\bm{\lambda}=
\begin{bmatrix}
\lambda_1,\ldots,\lambda_n
\end{bmatrix}'$,
what we are about to find in the end. We define the vector $\vec{I}_{\vec{X}}(\vec{t^*})=
\begin{bmatrix}
I_{\vec{X}}(\vec{t_1}),\ldots,I_{\vec{X}}(\vec{t_n})
\end{bmatrix}'$, 
using the calculated values from the previous step. So we get the equation system
\begin{equation}
\bm{\Psi}\bm{\lambda}=\vec{I}_{\vec{X}}. \label{er1}
\end{equation}
By solving \eqref{er1}, we can determine $\hat{\bm{\lambda}}$, however we run into some problems. First of all, $\hat{\bm{\lambda}}$ will most likely be a complex vector, which 
cannot be used for describing the distribution. The second problem is, that if the size of the grid is even, then the system \eqref{er1} is singular. That is because $\psi(-t;\alpha)=\overline{\psi(t;\alpha)}$ and $I_{\vec{X}}(\vec{-t})=\overline{I_{\vec{X}}(\vec{t})}$. Fortunately, we can deal with these problems using some modifications.

\subsubsection*{Step 4/1}
Let's assume that we try to find $\hat{\bm{\lambda}}$ based on even number of points ($n=2k$), where $\vec{t_j}=\vec{s_j}$ as before. We saw that using an even number of points is causing singularity in the system, but we can take advantage on this symmetry. In this case $\vec{I}_{\vec{X}}(\vec{t_i})=\overline{\vec{I}_{\vec{X}}(\vec{t_{i+k}})}$ and $\psi\left(\vec{t_i}^{\mathsf{T}}\vec{s_j};\alpha\right)=\overline{\psi\left(\vec{t_{i+k}}^{\mathsf{T}}\vec{s_j};\alpha\right)}$, as we have seen before. For these pairs it follows that 

\[\Re I_i=\dfrac{I_i+I_{i+k}}{2}=\sum_{j=1}^{n}\Re \psi_{i,j}\lambda_{j}\]

and 

\[\Im I_i=-\dfrac{I_i-I_{i+k}}{2}=\sum_{j=1}^{n}\Im \psi_{i,j}\lambda_{j},\] 

where $\vec{I}_{\vec{X}}(\vec{t_i})=I_i$ and $\psi_{i,j}=\psi\left(\vec{t_i}^{\mathsf{T}}\vec{s_j};\hat{\alpha}^*\right)$. 
We can now define a new $n\times 1$ vector with the real and imaginary parts of $I_{\vec{X}}(\vec{t})$ as 
\[\bm{c}=
\begin{bmatrix}
\Re I_1,\Re I_2,\ldots,\Re I_k,\Im I_1, \Im I_2,\ldots,\Im I_k
\end{bmatrix}\]
and a new $n\times n$ matrix $\bm{A}$ as
\[
a_{i,j}=
\begin{cases}
\Re \psi_{i,j}, \quad i=1,\ldots,k\\
\Im \psi_{i,j}, \quad i=k+1,\ldots,n
\end{cases}
\]
The system $\bm{A}\bm{\lambda}=\bm{c}$ is now non-singular and the solution will be a real vector, however still not usable. The problem is, that the solution may contain negative weights, which we cannot interpret.

\subsubsection*{Step 4/2}
To get non-negative weights we must modify the system once more. To be able to guarantee non-negativity, we redefine the problem as a quadratic programming problem as
\[
\min_{\bm{\lambda}} \norm{\bm{c}-\bm{A\lambda}}^2=\min_{\bm{\lambda}}  (\bm{c}-\bm{A\lambda})^{\mathsf{T}}(\bm{c}-\bm{A\lambda}),\quad \bm{\lambda}\ge 0.
\]
We estimate $\Lambda$ using this approach, implemented in \textsf{R} programming langauge.

\subsection{Properties}
Before going on, we need to note some important facts and properties about the estimation procedure.
\begin{itemize}
\item We can approximate the real spectral measure with a discrete $\Lambda$. Byczkowski, Nolan and Rajput showed in \cite{11}, that to a stable vector $\bm{X}$, with $\Lambda$ spectral measure, where $0<\alpha<2$ there exists a discrete $\Lambda^*$, such that
\[
\sup_{\bm{x}\in \mathbb{R}^{d}} \lvert p(\bm{x}) - p^*(\bm{x}) \rvert \leq \epsilon, 
\]
where $p(\bm{x})$ is the theoretical density, $p^*(\bm{x})$ is the corresponding density to $\Lambda^*$, $\epsilon>0$. 
Since both the quantile and the maximum likelihood methods are consistent and asymptotically unbiased, using the suggested approach to estimate parameters of the projections give consistent results in the proposed multivariate estimation procedure, we introduce in the next subsection.

\item The number of points of $S_2$ has to be an even number, $n=2k$, where $k\in \mathbb{N}\setminus \{1\}$. We saw in step 4/1, that we need this condition in order to be able to perform the necessary transformations. $k=1$  would result in a simple estimation of the first marginal distribution with the selected set of points of $S_2$. Apart from these, $n$ is a free parameter, but choosing $n$ as a power of two is the most preferable.
 Finding the appropriate number of points is not trivial. If the chosen $n$ is not large enough, the fitted distribution's dependence structure will not match the sample's. However, if $n$ is too large, the distribution can be overfitted, although theoretically it would give us the best results.

\end{itemize}

\subsection{Parameter estimation in higher dimensions}
In $d>2$ dimensions the estimation procedure gets a bit more difficult, because $\Lambda$ is concentrated on a sphere and not on a circle. The main difficulty is to select a  set of points from the surface of the sphere, for which we can repeat the same modifications as in Section \ref{estim}. We haven't found any papers dealing with estimation in higher dimensions, so the next generalization of the estimation, built on the previously seen bivariate method, is a new method, having a fast running time.

The key of estimating parameters for a $d>2$ dimensional stable distribution is to select points of $S_d$ pairwise from the marginals. Our suggestion is to choose the points from the circular cross section of the sphere, where 
all but two coordinates is always 0 for a given circle. This means, by selecting $n$ points 
for each circle, we will perform the estimation, based on ${d \choose 2} \cdot n$ points altogether.
\begin{figure}[H]\label{circ}
	\centering		
	\includegraphics[scale=0.35]{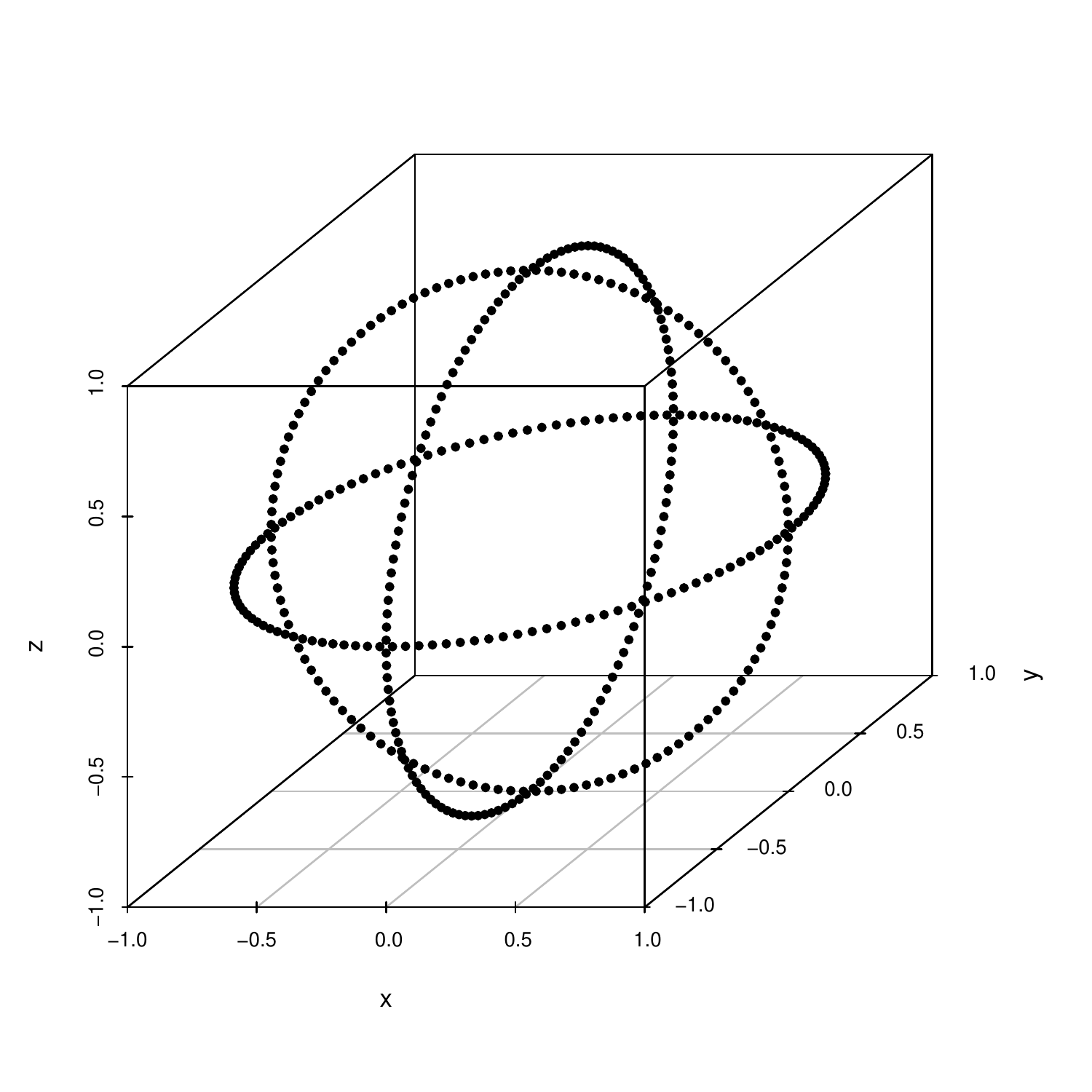}
	\caption{The suggested set of points from $S_3$}
\end{figure}

\subsubsection*{Step 1}
The first step is analogous to the bivariate estimation method. In this case, we have to estimate the vector $\bm{\delta}=\begin{bmatrix} \delta_1, \ldots , \delta_d \end{bmatrix}^{\mathsf{T}}$ by components with e.g. quantile method, which we have to subtract from the original sample to have the sample shifted to the origin.

\subsubsection*{Step 2}
We saw at the beginning of Section \ref{estim}, that the number of points from a circle had to be even. We still need this assumption, but for every individual circular cross section of the sphere. Additionally, we can't choose the same points from every circular cross section, because then we would be having duplicated points at the intersections of them and it would give us uninterpretable results. Therefore we choose the points rotated as
\[
\vec{s^{l,k}_j}=\left(0,\ldots,0,\underbrace{\cos\left(\frac{2\pi(j-1)}{n}+\frac{\pi}{n}\right)}_{\text { l-th coordinate}},0,\ldots,0,\underbrace{\sin\left(\frac{2\pi(j-1)}{n}+\frac{\pi}{n}\right)}_{\text { k-th coordinate}},0,\ldots,0\right),  
\]
where $\vec{s^{l,k}_j}$ is the circular cross section from the sphere constructed for the $l$-th and $k$-th marginals, $l\neq k$, $j=1,\ldots,n$. We pick the grid points as $\vec{t^{l,k}_j}=\vec{s^{l,k}_j}$ to be able to compute the projections
$\langle\vec{t^{l,k}_j},\vec{X_1}\rangle,\ldots,\langle\vec{t^{l,k}_j},\vec{X_m}\rangle$.

\subsubsection*{Step 3}
We have to calculate \eqref{proj1}, \eqref{proj2} and \eqref{proj3} for every projection as before. The pooled $\alpha$ remains essentially the same, the only real difference is that it is calculated from more projections as $\hat{\alpha}^*=\frac{1}{{d \choose 2} \cdot n}\sum_{l\neq k}\sum_{j=1}^n\hat{\alpha}(\vec{t^{l,k}_j})$. After these, we can compute the values
\[
\vec{I_{l,k}(t_j)}=\sum_{j=1}^n\psi\left(\vec{t}^{\mathsf{T}}\vec{s_j};\hat{\alpha}^*\right)\lambda^{l,k}_j,
\]
where $l\neq k$ and $j=1,\ldots,n$. However, the equation system we solved in Subsection \ref{estim} needs to be modified. 

\subsubsection*{Step 4}
The modified system is based on the matrices
\begin{align*}
\bm{\Psi^*}=
\begin{bmatrix}
\bm{\Psi_{1,2}} & \bm{0} & \ldots & \bm{0} \\
\bm{0} & \ddots &  & \vdots  \\
\vdots &  & \ddots & \bm{0}  \\
\bm{0} & \dots & \bm{0} & \bm{\Psi_{d-1,d}}
\end{bmatrix},
\quad
\vec{I^*}=
\begin{bmatrix}
\vec{I_{1,2}}\\
\vec{I_{1,3}}\\
\vdots\\
\vec{I_{d-1,d}}
\end{bmatrix},
\end{align*}
where $\bm{\Psi^*}\in\mathbb{R}^{{d \choose 2 }\cdot n \times{d \choose 2 }\cdot n}$, contains every calculated $\bm{\Psi_{l,k}}$ matrices, which are the same as \eqref{est_mtx}, but calculated from the $l$-th and $k$-th marginals. The $\bm{\Psi^*}$ matrix has the $\bm{\Psi_{l,k}}$ matrices in its diagonal, while its other elements are zero. The vector $\vec{I^*}\in\mathbb{R}^{{d \choose 2 }\cdot n}$ is modified with the same logic as $\bm{\Psi^*}$, so it contains the  vectors $\vec{I_{l,k}}$ combined together. Now, we are in the position to define the equation system
\begin{equation}
\bm{\Psi^*}\bm{\lambda^*}=\vec{I^*}_{\vec{X}}, \label{er2}
\end{equation}
which also has to be modified, because it is singular too due to the symmetrical construction of the points. 

\subsubsection*{Step 4/1}
We have to restrict the method to even number of points ($n=2r$) as before. Now, it is true, that $\vec{I}_{\vec{X}}(\vec{t^{l,k}_i})=\overline{\vec{I}_{\vec{X}}(\vec{t^{l,k}_{i+r}})}$ and $\psi\left(\vec{t_i}^{\mathsf{T}}\vec{s_j};\alpha\right)=\overline{\psi\left((\vec{t_{i+r}}^{{{\vec{l,k}}}})^{\mathsf{T}}\vec{s_j};\alpha\right)}$. We have to do the same transformation on the system as before in Section \ref{estim}, so we calculate the vectors

\[\Re I^{l,k}_i=\dfrac{I^{l,k}_i+I^{l,k}_{i+r}}{2}=\sum_{j=1}^{n}\Re \psi^{l,k}_{i,j}\lambda_{j}\] 

\[\Im I^{l,k}_i=-\dfrac{I^{l,k}_i-I^{l,k}_{i+r}}{2}=\sum_{j=1}^{n}\Im \psi^{l,k}_{i,j}\lambda_{j},\] 

where $\vec{I}_{\vec{X}}(\vec{t^{l,k}_i})=I^{l,k}_i$ and $\psi^{l,k}_{i,j}=\psi\left((\vec{t^{l,k}_i})^{\mathsf{T}}\vec{s_j};\hat{\alpha}^*\right)$. We now define the new ${d \choose 2 }\cdot n\times 1$ vector with the real and imaginary parts of $I_{\vec{X}}(\vec{t^{l,k}})$ as 
\[\bm{c^*}=
\begin{bmatrix}
\Re I^{1,2}_1,\Im I^{1,2}_1,\Re I^{1,2}_2,\Im I^{1,2}_2,\ldots,\Re I^{1,2}_r,\Im I^{1,2}_r,\Re I^{1,3}_1,\Im I^{1,3}_1,\ldots,\Re I^{n-1,n}_r,\Im I^{n-1,n}_r
\end{bmatrix}\]
and the new ${{d \choose 2 }\cdot n \times{d \choose 2 }\cdot n}$ matrix $\bm{A^*}$ as
\[
a^*_{i,j}=
\begin{cases}
\Re \psi^{1,2}_{i,j}, \quad i,j=1,\ldots,r\\
\Im \psi^{1,2}_{i,j}, \quad i,j=r+1,\ldots,n\\

\Re \psi^{1,3}_{i,j}, \quad i,j=n+1,\ldots,n+r\\
\Im \psi^{1,3}_{i,j}, \quad i,j=n+r+1,\ldots,2n\\
\vdots \\
\Re \psi^{d-1,d}_{i,j}, \quad i,j={d \choose 2}(n-1)+1,\ldots,{d \choose 2}(n-1)+r\\
\Im \psi^{d-1,d}_{i,j}, \quad i,j={d \choose 2}(n-1)+r+1,\ldots,{d \choose 2}n
\end{cases}
\]
Now the system $\bm{A^*}\bm{\lambda^*}=\bm{c^*}$ is non-singular and real, so we can perform the last modification step, to get positive weights.

\subsubsection*{Step 4/2}
We can analogously redefine the problem as a quadratic programming problem with $\vec{A^*}$ and $\vec{c^*}$:
\[
\min_{\bm{\lambda}} \norm{\bm{c^*}-\bm{A^*\lambda^*}}^2=\min_{\bm{\lambda^*}}(\bm{c^*}-\bm{A^*\lambda^*})^{\mathsf{T}}(\bm{c^*}-\bm{A^*\lambda^*}),\quad \lambda\ge 0.
\]
The solution gives us the desired results for $\bm{\lambda}$.

Due to the computational complexity of the method, we expect it to be applicable in its current form for moderately high dimensions only. In the next section we apply the estimation algorithm for a three-dimensional data set. 

\section{Applications }
In the applications, we fit multivariate stable distributions to cryptocurrency daily logreturns with large market capitalizations: Bitcoin, Ripple and Litecoin. 

The cryptocurrency data are from \url{www.kaggle.com}. These data and many more cryptocurrencies are also available in the recent \verb|crypto| package of \textsf{R}\cite{26}. The calculations of probabilities and sampling from univariate stable distribution were done with the help of the package \verb|stabdist|\cite{20}, while the univariate parameter estimations were performed using \verb|fBasics| \cite{21}. We used both of them for our own codes in the multivariate estimation and sampling, along with the package \verb|quadprog| \cite{22}, which solves the QP problem. Anderson-Darling tests were done by the package \verb|ADGofTest| \cite{23}, the determination of optimal block lengths for bootstrapping using \verb|ns| \cite{24} and the density estimations by \verb|ks| \cite{25}. 

\subsection{Data and the univariate estimation, goodness of fit} \label{app1}
The data is from April of 2013 to February of 2018.  The daily closing prices and the daily logreturns of the three assets can be seen on the figures \ref{cp} and \ref{lr} below.
\begin{figure}[H]\label{cp}
	\centering
		\includegraphics[scale=0.45]{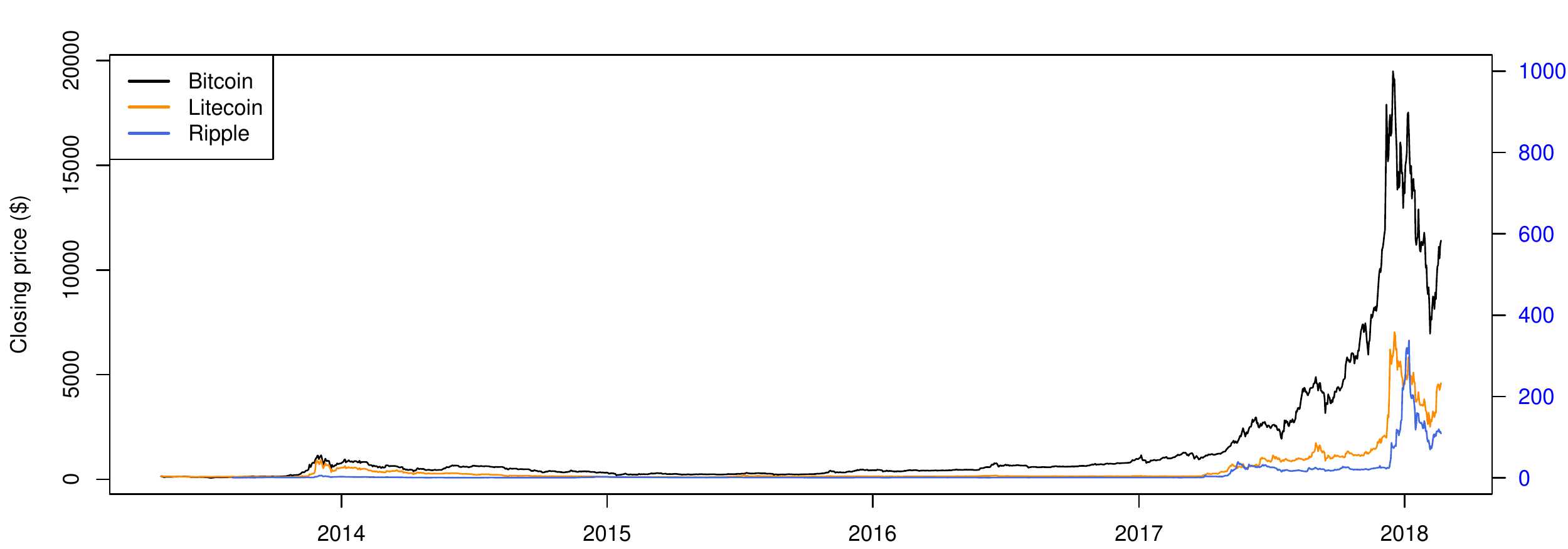}
		\caption{Prices of the selected cryptocurrencies in US dollar. Axis for Bitcoin is on the left, axis for Litecoin and Ripple on the right with different scale. The price of Ripple is multiplied by 100 for better visualization.}
\end{figure}

\begin{figure}[H]\label{lr}
	\centering
		\includegraphics[scale=0.45]{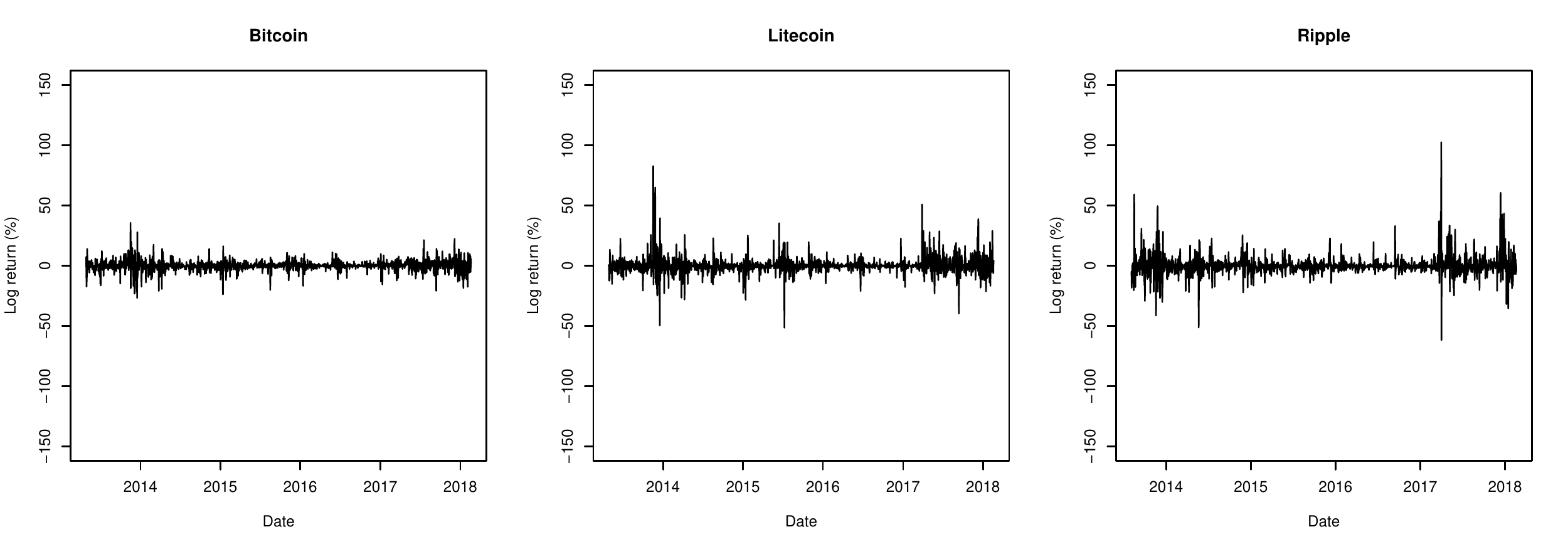}
		\caption{Daily logreturns of the selected cryptocurrencies}
\end{figure}

\begin{figure}[H]\label{gp}
	\centering
	\includegraphics[scale=0.35]{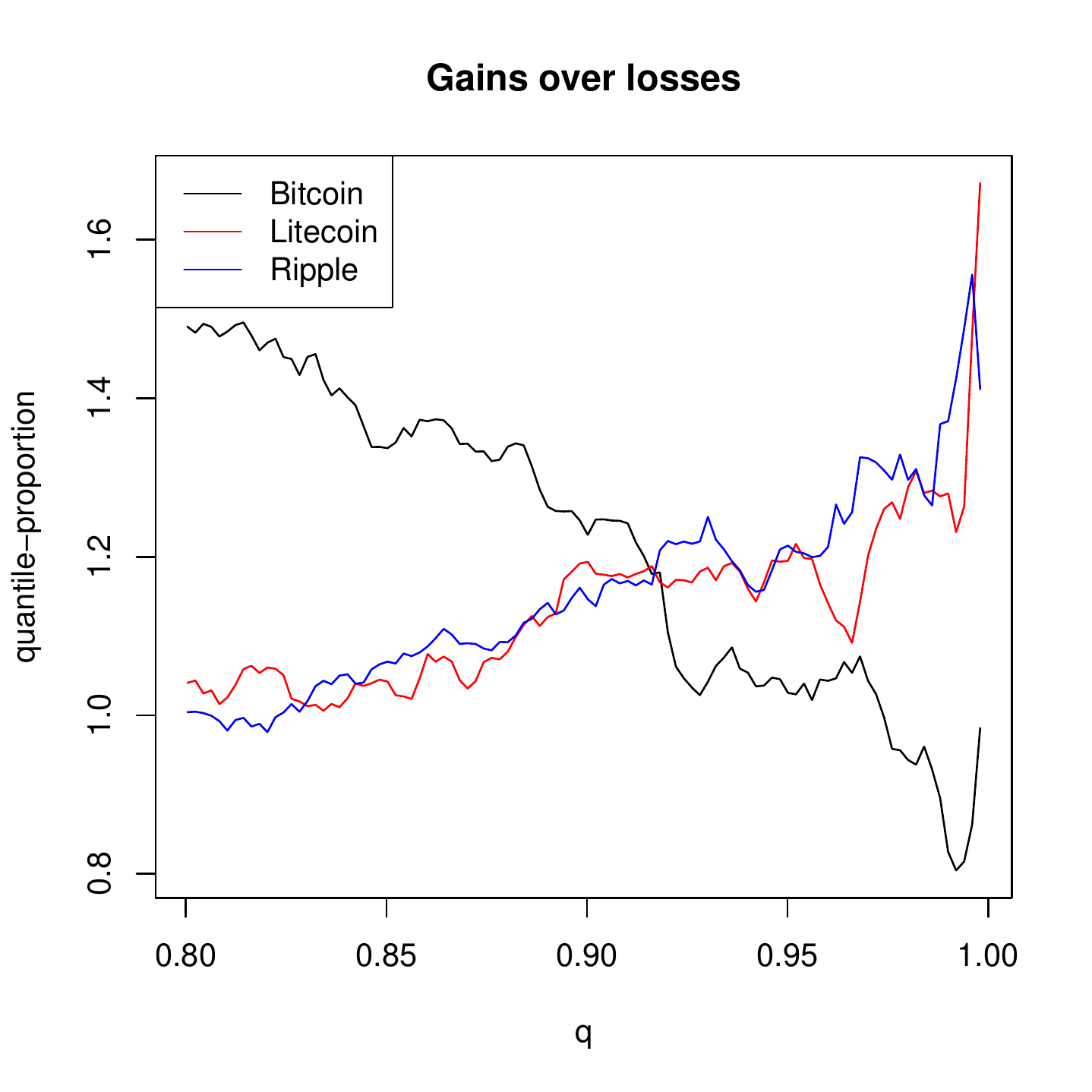}
	\caption{Comparison of the high and low $q-$quantiles of the daily logreturns of the selected cryptocurrencies.}
\end{figure}

To be able to apply multivariate stable distributions to the logreturns, we have to check whether the marginal distributions of them can be accepted as being stable at all. So before carrying out the analysis, we checked whether the losses and the gains are of the same magnitude (needed for a stable distribution). 
As a graphical tool, the high quantiles were compared on Figure \ref{gp}, showing that in two cases the high quantiles for the gains were even larger (values over 1) than those for the losses. For traditional stock one would expect values (possibly substantially) below 1.

We split the whole data into 10 overlapping, $\approx 40$ months wide windows with $\approx 1000$ observations. The first period is the oldest, and the 10th period is the most recent in the applications. 
After this initial step, we estimated the parameters using MLE for $\alpha$ and quantile method for $\beta,\gamma$ and $\delta$.

\begin{table}[H]
\centering
\begin{tabular}{|l|l|l|l|l|l|l|l|l|}
\cline{1-4} \cline{6-9}
\multicolumn{4}{|c|}{Bitcoin parameters} &  & \multicolumn{4}{c|}{Litecoin parameters} \\ \cline{1-4} \cline{6-9} 
\multicolumn{1}{|c|}{$\alpha$} & \multicolumn{1}{c|}{$\beta$} & \multicolumn{1}{c|}{$\gamma$} & \multicolumn{1}{c|}{$\delta$} &  & \multicolumn{1}{c|}{$\alpha$} & \multicolumn{1}{c|}{$\beta$} & \multicolumn{1}{c|}{$\gamma$} & \multicolumn{1}{c|}{$\delta$} \\ \cline{1-4} \cline{6-9} 
\cellcolor[HTML]{EFEFEF}1.319 & -0.014 & \cellcolor[HTML]{EFEFEF}1.768 & 0.064 &  & \cellcolor[HTML]{EFEFEF}1.211 & 0.015 & \cellcolor[HTML]{EFEFEF}2.058 & -0.246 \\ \cline{1-4} \cline{6-9} 
\cellcolor[HTML]{EFEFEF}1.292 & 0.043 & \cellcolor[HTML]{EFEFEF}1.636 & 0.033 &  & \cellcolor[HTML]{EFEFEF}1.15 & 0.064 & \cellcolor[HTML]{EFEFEF}1.824 & -0.24 \\ \cline{1-4} \cline{6-9} 
\cellcolor[HTML]{EFEFEF}1.254 & -0.014 & \cellcolor[HTML]{EFEFEF}1.418 & 0.072 &  & \cellcolor[HTML]{EFEFEF}1.117 & 0.016 & \cellcolor[HTML]{EFEFEF}1.607 & -0.121 \\ \cline{1-4} \cline{6-9} 
\cellcolor[HTML]{EFEFEF}1.291 & -0.067 & \cellcolor[HTML]{EFEFEF}1.243 & 0.069 &  & \cellcolor[HTML]{EFEFEF}1.15 & -0.041 & \cellcolor[HTML]{EFEFEF}1.426 & 0.021 \\ \cline{1-4} \cline{6-9} 
\cellcolor[HTML]{EFEFEF}1.3 & 0.014 & \cellcolor[HTML]{EFEFEF}1.131 & 0.08 &  & \cellcolor[HTML]{EFEFEF}1.144 & -0.035 & \cellcolor[HTML]{EFEFEF}1.21 & 0.016 \\ \cline{1-4} \cline{6-9} 
\cellcolor[HTML]{EFEFEF}1.295 & -0.059 & \cellcolor[HTML]{EFEFEF}1.144 & 0.163 &  & \cellcolor[HTML]{EFEFEF}1.107 & -0.027 & \cellcolor[HTML]{EFEFEF}1.169 & 0.012 \\ \cline{1-4} \cline{6-9} 
\cellcolor[HTML]{EFEFEF}1.27 & -0.005 & \cellcolor[HTML]{EFEFEF}1.136 & 0.217 &  & \cellcolor[HTML]{EFEFEF}1.063 & 0.046 & \cellcolor[HTML]{EFEFEF}1.175 & -0.02 \\ \cline{1-4} \cline{6-9} 
\cellcolor[HTML]{EFEFEF}1.273 & 0.001 & \cellcolor[HTML]{EFEFEF}1.167 & 0.244 &  & \cellcolor[HTML]{EFEFEF}1.058 & 0.115 & \cellcolor[HTML]{EFEFEF}1.262 & -0.054 \\ \cline{1-4} \cline{6-9} 
\cellcolor[HTML]{EFEFEF}1.225 & -0.054 & \cellcolor[HTML]{EFEFEF}1.223 & 0.283 &  & \cellcolor[HTML]{EFEFEF}1.043 & 0.149 & \cellcolor[HTML]{EFEFEF}1.395 & -0.078 \\ \cline{1-4} \cline{6-9} 
\cellcolor[HTML]{EFEFEF}1.184 & -0.07 & \cellcolor[HTML]{EFEFEF}1.381 & 0.361 &  & \cellcolor[HTML]{EFEFEF}1.051 & 0.096 & \cellcolor[HTML]{EFEFEF}1.5 & -0.054 \\ \cline{1-4} \cline{6-9} 

\end{tabular}
\caption{The estimated parameters for Bitcoin and Litecoin daily logreturns for every period (from oldest (1.) to the newest (10.)). Note that with time, $\alpha$ decreases, while $\gamma$ slightly grows after a rapid fall. The effects of these parameter changes are that more probability is getting concentrated on the tails. Also, in the case of Litecoin, a slight positive skewness appears in the last periods.}
\label{T1}
\end{table}

As a goodness-of-fit procedure, we performed Anderson-Darling test on Bitcoin and Litecoin logreturns for all periods (using simulated critical values, as the effect of parameter estimation is distribution-dependent).

\begin{table}[H]
\centering
\resizebox{\textwidth}{!}{%
\begin{tabular}{lllllllllll}
\hline

\multicolumn{1}{|l|}{Bitcoin} & \multicolumn{1}{c|}{1.} & \multicolumn{1}{c|}{2.} & \multicolumn{1}{c|}{3.} & \multicolumn{1}{c|}{4.} & \multicolumn{1}{c|}{5.} & \multicolumn{1}{c|}{6.} & \multicolumn{1}{c|}{7.} & \multicolumn{1}{c|}{8.} & \multicolumn{1}{c|}{9.} & \multicolumn{1}{c|}{10.} \\ \hline

\multicolumn{1}{|l|}{Critical Value} 	 & \multicolumn{1}{l|}{2.428}	 & \multicolumn{1}{l|}{2.387} 	 & \multicolumn{1}{l|}{2.399} 	 & \multicolumn{1}{l|}{2.664} 	 & \multicolumn{1}{l|}{2.343} 	 & \multicolumn{1}{l|}{2.426} 	 & \multicolumn{1}{l|}{2.55} 	 & \multicolumn{1}{l|}{2.659} 	 & \multicolumn{1}{l|}{2.57} 	 & \multicolumn{1}{l|}{2.491}  \\ \hline
										
\multicolumn{1}{|l|}{Test statistic} 	 & \multicolumn{1}{l|}{1.709}	 & \multicolumn{1}{l|}{1.123} 	 & \multicolumn{1}{l|}{0.824} 	 & \multicolumn{1}{l|}{1.232} 	 & \multicolumn{1}{l|}{1.269} 	 & \multicolumn{1}{l|}{1.582} 	 & \multicolumn{1}{l|}{1.777} 	 & \multicolumn{1}{l|}{2.473} 	 & \multicolumn{1}{l|}{\cellcolor[HTML]{CB0000}3.906} 	 & \multicolumn{1}{l|}{\cellcolor[HTML]{CB0000}4.054}  \\ \hline

 &  &  &  &  &  &  &  &  &  &  \\ \hline

\multicolumn{1}{|l|}{Litecoin} & \multicolumn{1}{c|}{1.} & \multicolumn{1}{c|}{2.} & \multicolumn{1}{c|}{3.} & \multicolumn{1}{c|}{4.} & \multicolumn{1}{c|}{5.} & \multicolumn{1}{c|}{6.} & \multicolumn{1}{c|}{7.} & \multicolumn{1}{c|}{8.} & \multicolumn{1}{c|}{9.} & \multicolumn{1}{c|}{10.} \\ \hline

 \multicolumn{1}{|l|}{Critical Value} 	 & \multicolumn{1}{l|}{2.625}	 & \multicolumn{1}{l|}{2.591} 	 & \multicolumn{1}{l|}{2.334} 	 & \multicolumn{1}{l|}{2.577} 	 & \multicolumn{1}{l|}{2.358} 	 & \multicolumn{1}{l|}{2.366} 	 & \multicolumn{1}{l|}{2.594} 	 & \multicolumn{1}{l|}{2.548} 	 & \multicolumn{1}{l|}{2.172} 	 & \multicolumn{1}{l|}{1.967} \\ \hline			
													
 \multicolumn{1}{|l|}{Test statistic} 	 & \multicolumn{1}{l|}{0.747}	 & \multicolumn{1}{l|}{0.787} 	 & \multicolumn{1}{l|}{1.041} 	 & \multicolumn{1}{l|}{1.807} 	 & \multicolumn{1}{l|}{1.963} 	 & \multicolumn{1}{l|}{1.069} 	 & \multicolumn{1}{l|}{1.011} 	 & \multicolumn{1}{l|}{1.293} 	 & \multicolumn{1}{l|}{1.397} 	 & \multicolumn{1}{l|}{1.663} \\ \hline

\end{tabular}%
}
\caption{Results of the performed Anderson-Darling tests for every period. Rejection of the null hypothesis (univariate stability) happened only at Bitcoin's last two tested periods.}
\label{t3}
\end{table}

\begin{figure}[H]
	\centering
	\includegraphics[scale=0.35]{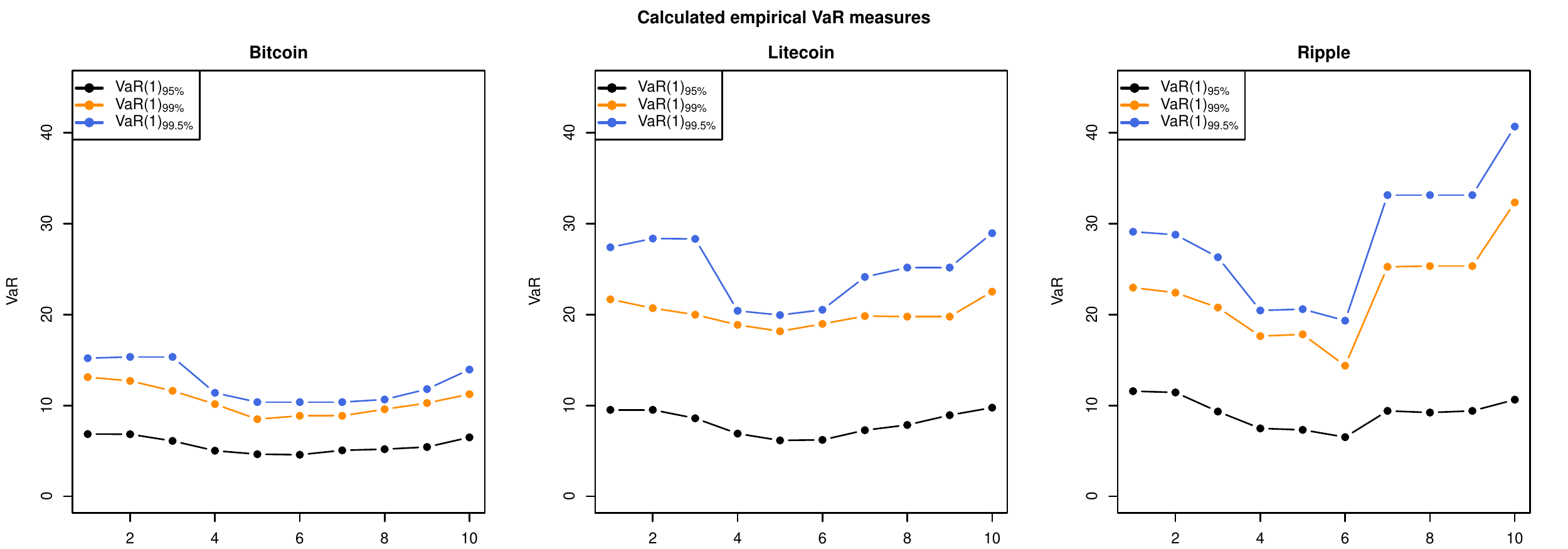}
	\caption{VaR estimators, based on the empirical distribution of the cryptocurrencies.}
	\label{ev}
	
\end{figure}

\begin{figure}[H]
	\centering
	\includegraphics[scale=0.35]{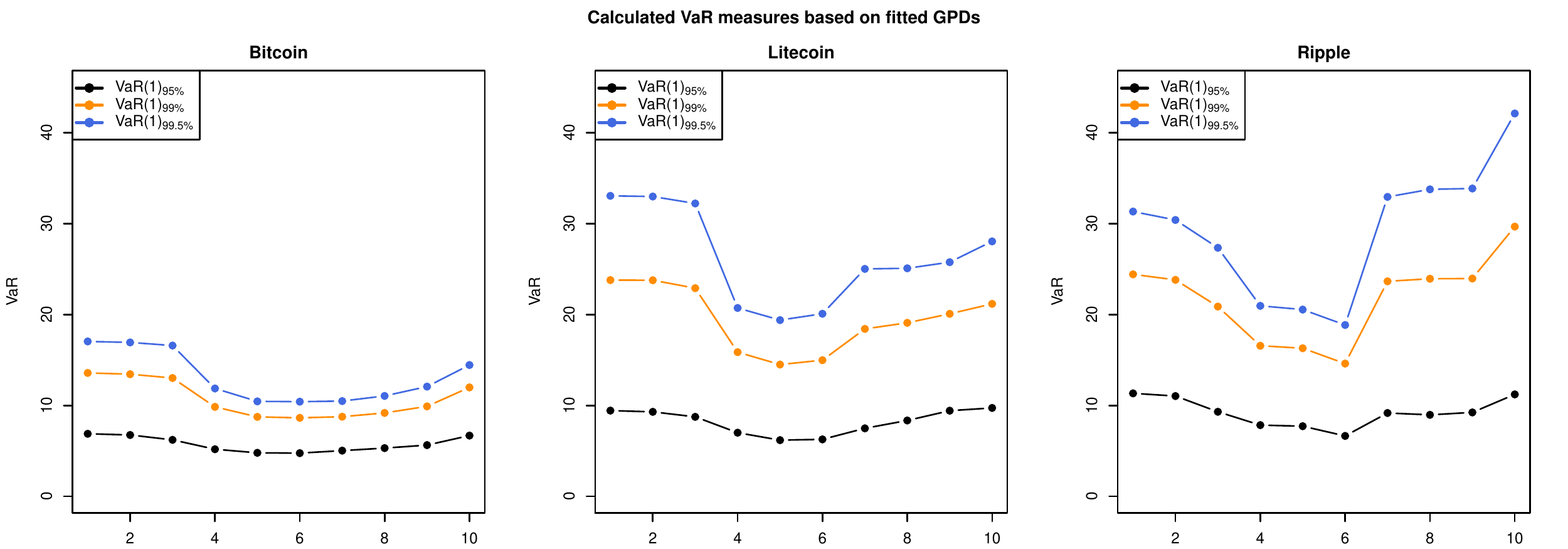}
	\caption{VaR estimators, based on the GPD model of the cryptocurrencies.}
	\label{gv}
	
\end{figure}

\begin{figure}[H]
	\centering
	\includegraphics[scale=0.35]{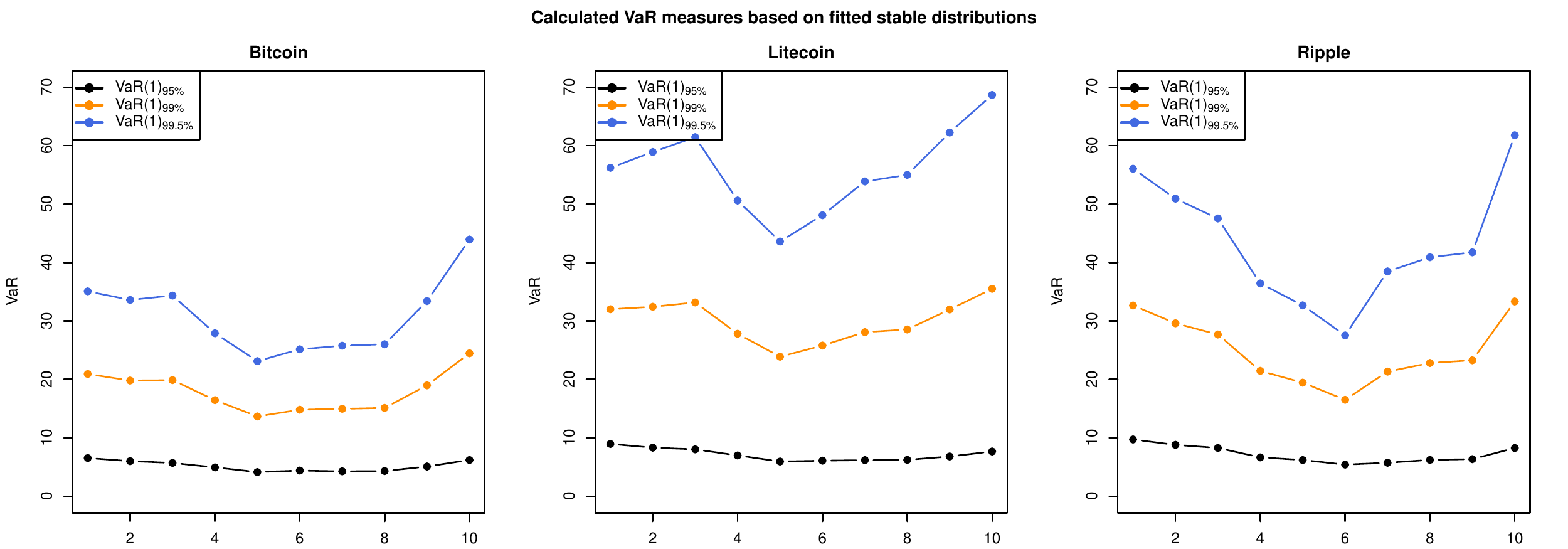}
	\caption{VaR estimators, based on the estimated stable distribution.}
	\label{sv}
	
\end{figure}

The results from the test statistics are promising, nevertheless we got two rejections. This may be caused by the rapid change of volatility over the last years, which is contained in the last two windows. The price fluctuated from a few hundred to over 10 thousand with a visible gain/loss asymmetry in these years. We still fit multivariate stable distributions to these two cases as well as it may show interesting changes in dependence structure between the two currencies. But before turning to the multivariate modelling, let us compare one of the most important risk measures: the VaR (value-at risk for the losses) for the investigated periods-calculated by three different methods: the empirical distribution (not recommended in practical use, due to the large variance of the high quantiles - and besides, it cannot produce higher quantile estimators than the actual observations, Figure \ref{ev}), the quite common peaks-over threshold approach (with the 0.97-quantile \% as the threshold,  Figure \ref{gv}) and the stable model,  Figure \ref{sv}.

We can observe that there is a less volatile period in the middle of the investigated time interval. And it is clear that the stable model gives higher quantiles if we estimate extreme high values - which may be considered as a more cautious approach.

\subsection{Multivariate estimation and goodness of fit}
\subsubsection{Bivariate model}
The estimation is performed using both ML and quantile method: the former is used for $\alpha$ at the two marginal distributions, and the quantile method for the other parameters on every projection, similarly as we can see in Table \ref{T1}. 
The pooled $\alpha^*$ is now calculated only from the estimated $\alpha$ parameters of the marginals. The main reason behind this, is that ML method usually performs better at estimating $\alpha$ (the quantile method tends to overestimate the tails by underestimating $\alpha$). This can be critical, since a small difference in the value of $\alpha$ means significant change in the probability of extremal events. Also, the other problem is, if we would like to estimate $\alpha$ 
for all the projections, we should definitely be using the quantile method, 
because even for one dataset, with $\approx 1000$ elements ML runs for at least 5 minutes. This amount of time is acceptable for two marginals, but it is not practical if we would like to use it for every projection. Computing the necessary values for the multivariate parameter estimation with these changes gives us the best results with the least consumed time.

\begin{table}[H]
\centering
\begin{tabular}{c|c|c|c|c|c|c|}
\cline{2-7}
\multicolumn{1}{l|}{} & \multicolumn{2}{c|}{8 points} & \multicolumn{2}{c|}{16 points} & \multicolumn{2}{c|}{32 points} \\ \hline
\multicolumn{1}{|c|}{Window} & \begin{tabular}[c]{@{}c@{}}Test\\ statistic\end{tabular} & \begin{tabular}[c]{@{}c@{}}Critical\\ value\end{tabular} & \begin{tabular}[c]{@{}c@{}}Test\\ statistic\end{tabular} & \begin{tabular}[c]{@{}c@{}}Critical\\ value\end{tabular} & \begin{tabular}[c]{@{}c@{}}Test\\ statistic\end{tabular} & \begin{tabular}[c]{@{}c@{}}Critical\\ value\end{tabular} \\ \hline
\multicolumn{1}{|c|}{1.} &	 \cellcolor[HTML]{CB0000}1.257 &	0.355	& 0.081	& 0.213	& 0.074	& 0.610 \\ \hline
\multicolumn{1}{|c|}{2.} &	1.390 &	2.342 &	0.169 &	1.115	& 0.049	& 0.221 \\ \hline
\multicolumn{1}{|c|}{3.} &	 \cellcolor[HTML]{CB0000}1.785 &	0.376	& \cellcolor[HTML]{CB0000}0.929 &	0.302	& 0.351	& 0.635 \\ \hline
\multicolumn{1}{|c|}{4.} &	 \cellcolor[HTML]{CB0000}2.434 &	0.563 &	0.276 &	1.066 &	0.525	& 0.682 \\ \hline
\multicolumn{1}{|c|}{5.} &	 \cellcolor[HTML]{CB0000}2.832 &	0.684	& \cellcolor[HTML]{CB0000}0.368 &	0.278	& 0.441	& 0.684 \\ \hline
\multicolumn{1}{|c|}{6.} &	 \cellcolor[HTML]{CB0000}2.518 &	0.495	& 0.481	& 0.495 &	0.423 &	0.444 \\ \hline
\multicolumn{1}{|c|}{7.} &	2.121 &	4.093	& \cellcolor[HTML]{CB0000}0.455 &	0.399	& \cellcolor[HTML]{CB0000}0.286 &	 0.230 \\ \hline
\multicolumn{1}{|c|}{8.} & 0.472 &	0.986	& 0.171	& 0.323 &	0.306	& 0.873 \\ \hline
\multicolumn{1}{|c|}{9.} &	2.540 &	5.315 &	0.361 &	1.141 &	0.214	&  2.810 \\ \hline
\multicolumn{1}{|c|}{10.} &	 \cellcolor[HTML]{CB0000}6.072 &	0.457	& \cellcolor[HTML]{CB0000}1.266 &	0.840	& \cellcolor[HTML]{CB0000}1.280 	& 0.335 \\ \hline

\end{tabular}
\caption{Results of the performed 
	Cramér--von Mises type  test statistics based on Kendall functions for the 3 different estimation approaches. Critical values were chosen at the 95\% significance level. The values with red background are the tests, where the null hypothesis had to be rejected.}
\label{kendallstat}
\end{table}
For the last period, none of the fitted multivariate stable distributions are acceptable. This is not that surprising, as we could see, that for the last period we couldn't fit an univariate stable distribution to Bitcoin's logreturns. For the second, eighth and ninth period, we didn't get rejection, however the Anderson-Darling test for Bitcoin at the ninth period resulted in a rejection. For the rest of the periods, apart from the seventh period, by increasing the number of points on the circle, we usually get acceptable fit. Also, it is important that some of the estimated critical values were extremal compared to the others. This could have been mitigated, if the number of simulations was higher, although most of these extremal values can be seen for one period (9.), so this may be partially caused by the underlying distribution.

\begin{figure}[H]
	\centering
	\includegraphics[scale=0.35]{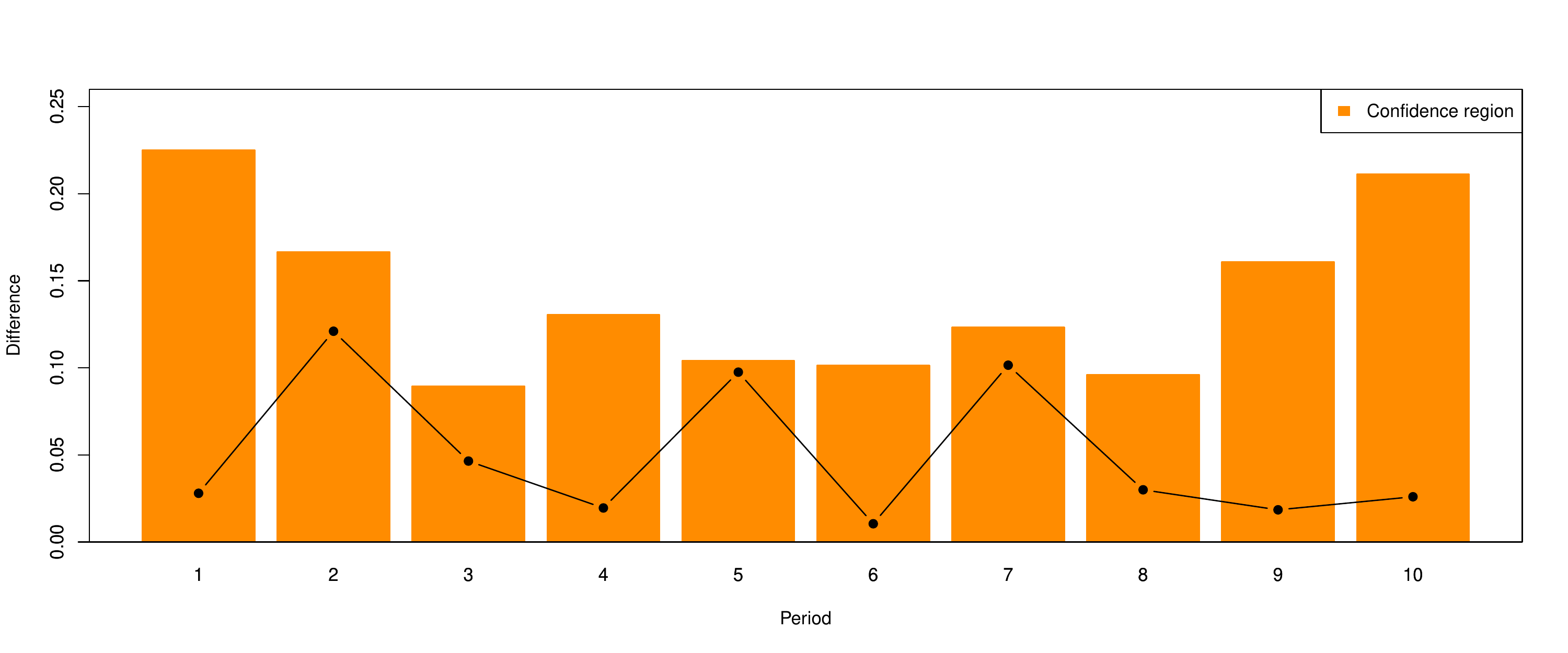}
	\caption{The result of the bootstrap-test for the equality of the $\alpha$ values for the coordinates}
	\label{alp}
\end{figure}

One more test is important: the equity of the $\alpha$ parameters of the two coordinates. There are no formal tests available for this purpose, but one can easily construct a bootstrap test by estimating the joint $\alpha$ value and then simulating $n$-element samples from this distribution. The marginal $\alpha$ values are estimated for these samples and the empirical 95\% upper confidence bound can be used as a critical value. Figure \ref{alp} shows these results, and we see that there were no rejections during the investigated periods. 

To be able to visualize and understand better the estimated spectral measures, it can be useful to look at the figures below. These plots are giving an idea about the shape of the density, characterized by the spectral measure.

\begin{figure}[H]
	\centering
		\includegraphics[scale=0.35]{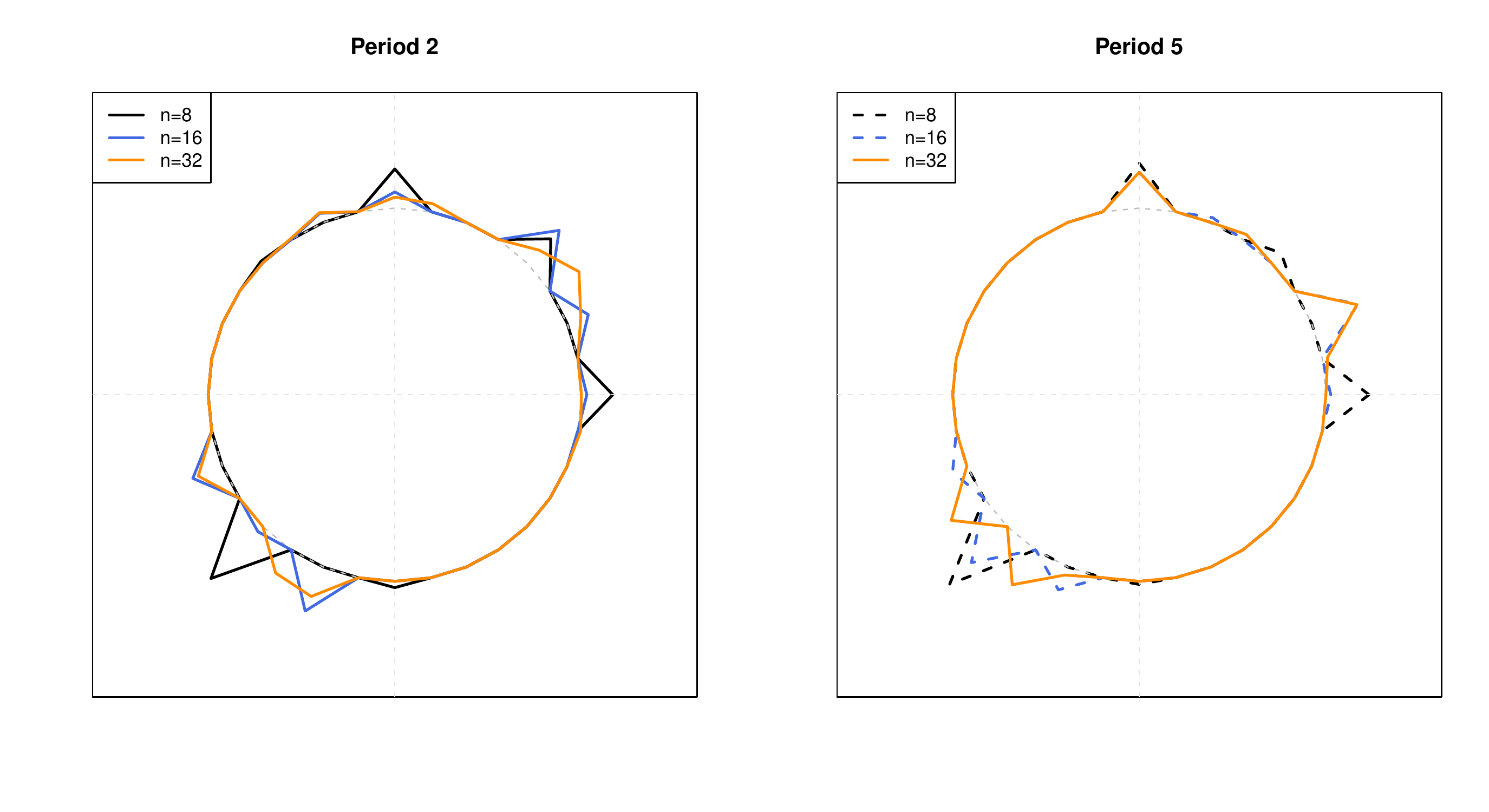}
\end{figure}

\vspace{-20mm}

\begin{figure}[H]
	\centering
		\includegraphics[scale=0.35]{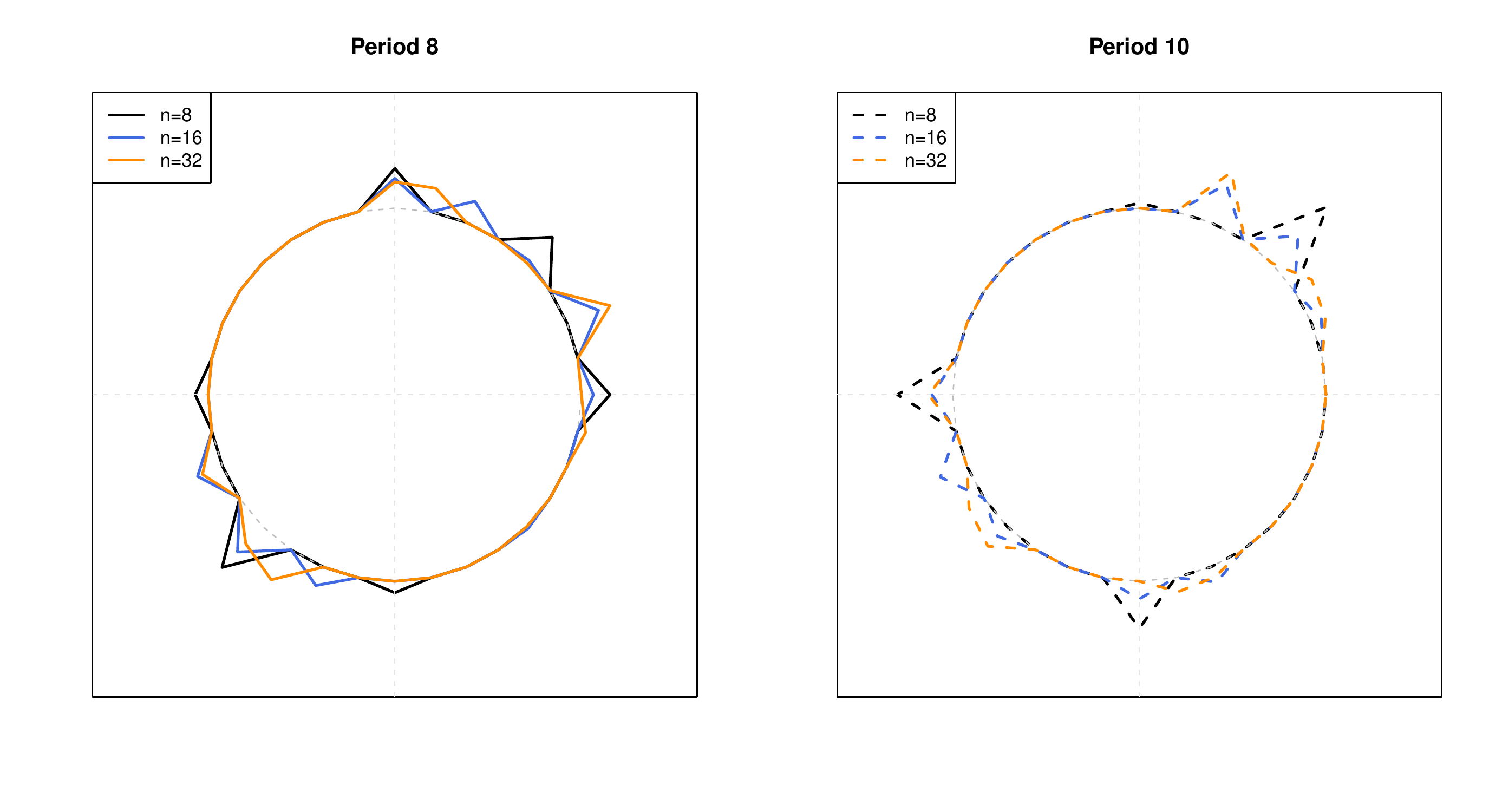}
				\caption{Visualization of the estimated spectral measures for the periods 2-5-8-10, with different number of points. Dashed line means that the fit was not acceptable (parallel to Table \ref{kendallstat}). 
					}
\end{figure}

The different number of points are giving significantly different estimated spectral measures. With less points from $S_2$, the measure is concentrated on fewer points and results in a simpler dependence structure. Best example is the measure at period 2, where all the fits are acceptable. It is visible, that the weights in the lower left quadrant 
spread to more and more points when increasing the number of points. Another good example, is period 5, where with 8 points, the procedure found positive weight on the very first point $s_1$, but with 32 points, the method couldn't find any.

We have got the best results, when the estimation was done on 32 points, which is expected, based on theoretical property showed in \cite{11}. However, in applications, especially in financial applications with such large number of points we can get overfitted models. This is absolutely true for cryptocurrencies, where the dependence structure can change really fast, due to their unpredictable nature.

We can say that the dependence structure changed over time, but the best way to visualise this is by the densities. Although, it would take a lot of time to compute the actual density function, we can simulate a larger sample from the given distribution and run a simple density estimation on it. The results of these for every period can be seen on the figures below.

\begin{figure}[H]
	\centering		
		\includegraphics[scale=0.35]{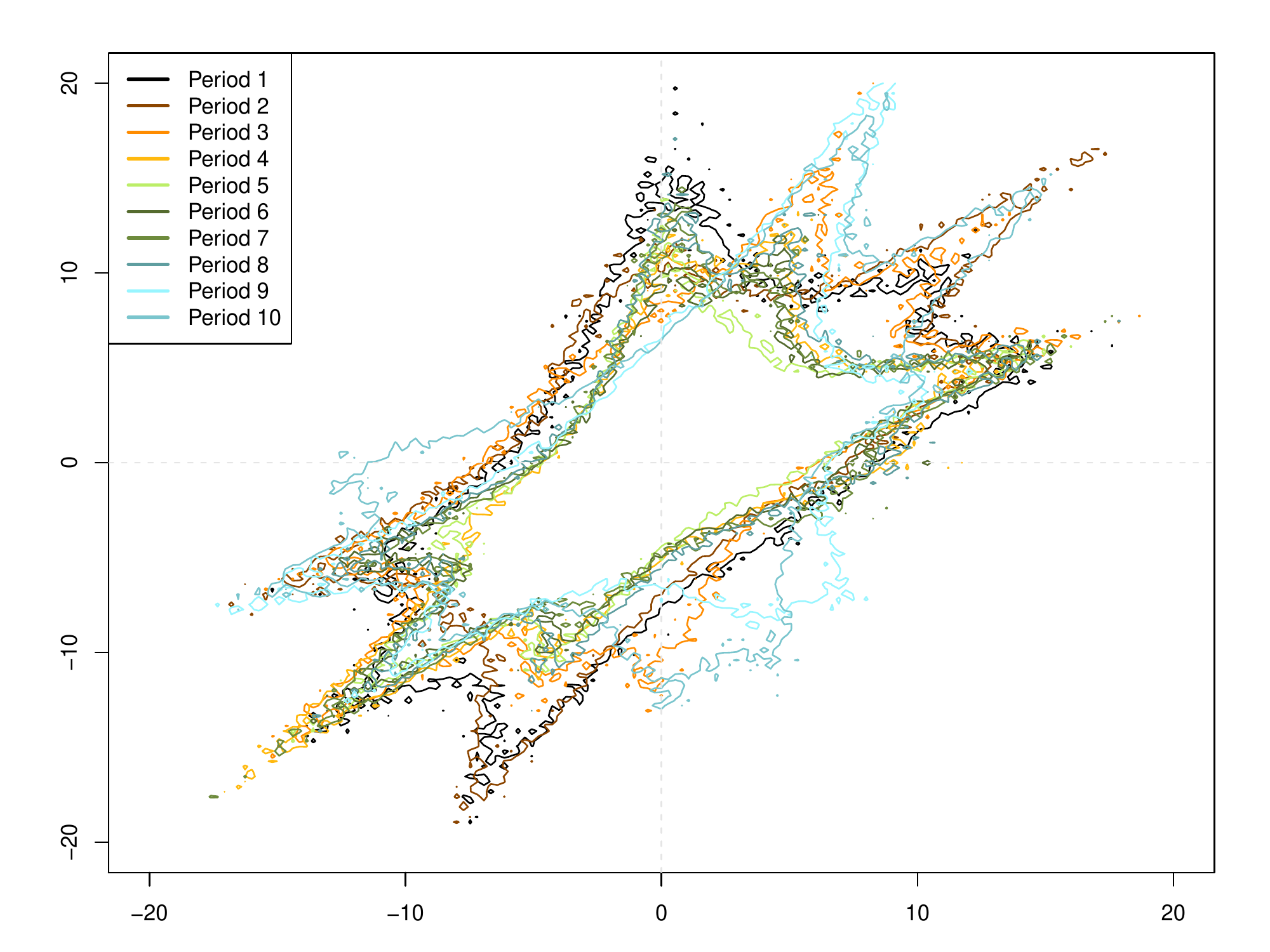}
\end{figure}

\begin{figure}[H]
	\centering		
		\includegraphics[scale=0.5]{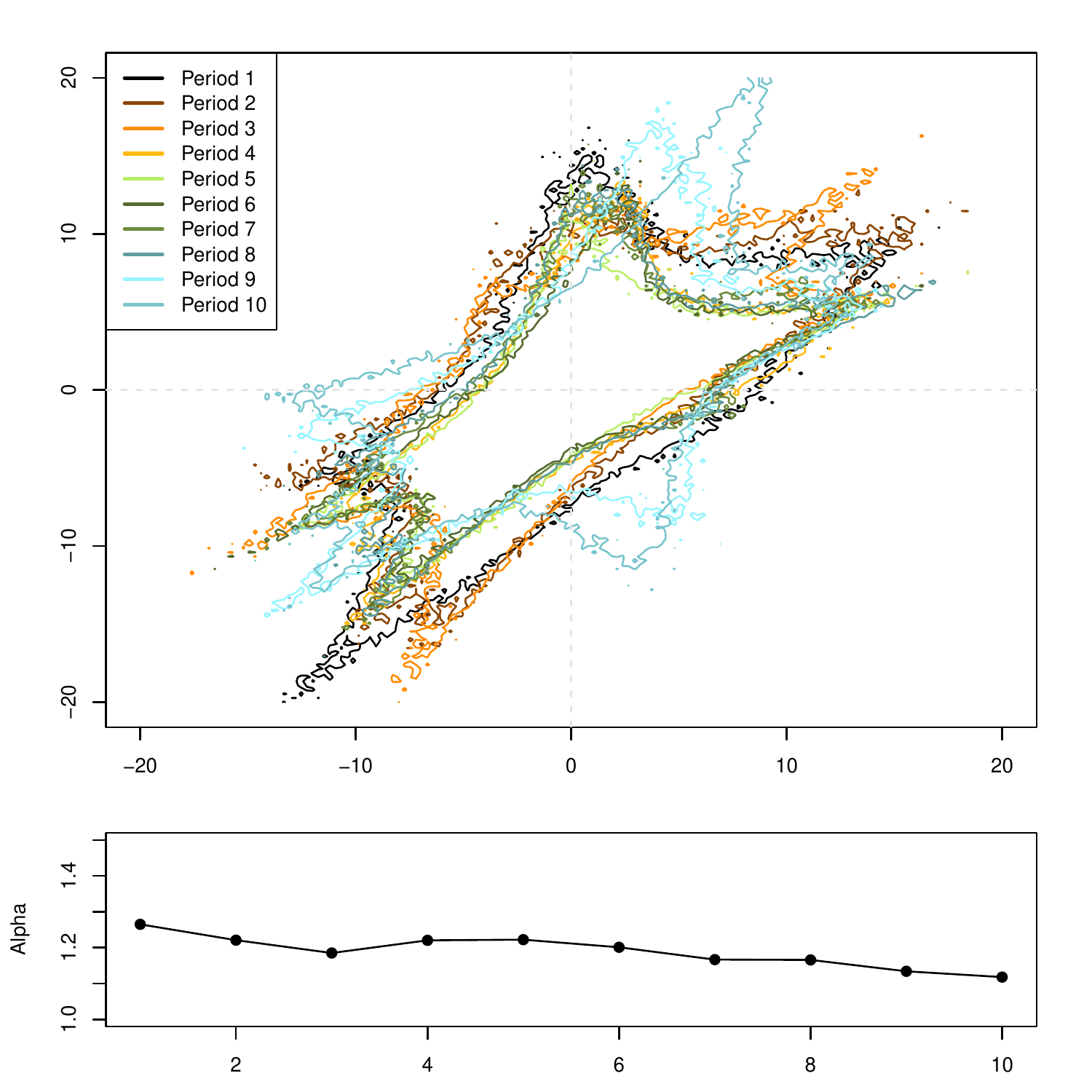}
		\caption{The 95\% probability covering regions based on the density estimations with the estimations on 16 points (first) and 32 points (second) from $S_2$, and the estimated pooled $\alpha^*$ for all periods. Distribution of Bitcoin and Litecoin logreturns on the horizontal and vertical axes in order.}
		\label{dens32}
\end{figure}

The dependence structure visibly changes in time for both calibrations. The two results are similar, the angles with larger weights are mostly present at both. The two dominant angles at the upper right quadrant are shifting with time, at first far from each other then in the end closer together, while moving back and forth in the interim periods. The lower left quadrant also has two dominant angles, both moving back and forth with time. In the last period both changes, having shifted these angles closer to the horizontal axis and a new angle appears in the lower right quadrant, giving more probability to opposite movement in price changes.
The contour lines around the covering region are nothing like the classic elliptic contour lines we are used to, e.g. the normal distribution. This is partially caused by the low estimated $\alpha$, giving heavy tails to the distribution and making the already dominant angles more dominant and spreading the covering region into a larger area.

We can easily calculate probabilities or risk measures with the distribution. We calculate probabilities, because a simple VaR should be calculated from the sum of the variables and the goal is to take into consideration the dependence structure directly. For comparison, we calculated the conditional probability 
\[
P(\text{Litecoin logreturn}<-10\% \mid \text{Bitcoin logreturn}<-10\%)
\]
based on the estimated stable distributions on 32 points and fitted bivariate normal distributions.
\begin{table}[H]
\centering
\resizebox{\textwidth}{!}{%
\begin{tabular}{|l|l|l|l|l|l|l|l|l|l|l|}
\hline
Period & \multicolumn{1}{c|}{1.} & \multicolumn{1}{c|}{2.} & \multicolumn{1}{c|}{3.} & \multicolumn{1}{c|}{4.} & \multicolumn{1}{c|}{5.} & \multicolumn{1}{c|}{6.} & \multicolumn{1}{c|}{7.} & \multicolumn{1}{c|}{8.} & \multicolumn{1}{c|}{9.} & \multicolumn{1}{c|}{10.} \\ \hline
Normal & 0.88\% & 0.31\% & 0.17\% & 0.25\% & 0.29\% & 0.23\% & 0.61\% & 2.95\% & 3.19\% & 3.51\% \\ \hline
Stable  & 46.5\% & 59.02\%& 76.81\%& 73.68\%& 72.91\%& 80\%& 89.84\%& 61.96\%& 66.98\% & 71.76\% \\ \hline
\end{tabular}%
}
\caption{Calculated conditional probabilities. The probabilities from the stable distribution are calculated from samples with the given parameters.}
\label{prob1}
\end{table}

The results are very illustrative. The conditional probabilities calculated from the fitted normal distributions are always around a few percent, but the probabilities from the fitted stable distributions are huge compared to them. These results are in line with the extremal (tail) independence and dependence of the normal and stable distributions, respectively. We say that a distribution has extremal independence or dependence, if the probabilities
\begin{align*}
\theta_l=\lim_{q\to 0} P(Y< F_2^{-1}(q)|X< F_1^{-1}(q)),\\
\theta_u=\lim_{q\to 1} P(Y\ge F_2^{-1}(q)|X\ge F_1^{-1}(q))
\end{align*}
tend to zero or nonzero, where $F_1$ and $F_2$ are the distribution functions of $X$ and $Y$, respectively \cite{18}. Although -10\% isn't an extreme quantile, these properties can already be observed from the calculations.

\subsubsection{Application in 3 dimensions}
Now we are able to look into fitting 3 dimensional stable distributions to the data. We fit the distribution to all the three cryptocurrency's logreturns that we showed in Section \ref{app1}. 

In this case, we fit only for the last three periods, which are the equivalent to periods 8, 9 and 10 from Section \ref{app1}. We have to keep in mind, that the AD test resulted in rejection for Bitcoin's last two period and the Kendall function based test completely rejected the last period for the three calibration. Despite these problems, these periods are the most interesting for us, because the drastic changes in the prices were observed in these periods. The estimation and testing is done with the same logic as before, therefore we estimate $\alpha$ with MLE, $\beta,\gamma$ and $\delta$ with the quantile method.
\begin{table}[H]
\centering
\begin{tabular}{|l|l|l|l|}
\hline
\multicolumn{4}{|c|}{Ripple parameters}                                                                                                                                \\ \hline
\multicolumn{1}{|c|}{$\alpha$}                     & \multicolumn{1}{c|}{$\beta$} & \multicolumn{1}{c|}{$\gamma$}                      & \multicolumn{1}{c|}{$\delta$} \\ \hline
\cellcolor[HTML]{EFEFEF}1.168 & 0.212                        & \cellcolor[HTML]{EFEFEF}1.565 & -0.479                        \\ \hline
\cellcolor[HTML]{EFEFEF}1.168 & 0.223                        & \cellcolor[HTML]{EFEFEF}1.619 & -0.485                        \\ \hline
\cellcolor[HTML]{EFEFEF}1.114 & 0.177                        & \cellcolor[HTML]{EFEFEF}1.882 & -0.473                        \\ \hline
\end{tabular}
\caption{The estimated parameters of the fitted stable distribution calculated from Ripple logreturns for the three periods}
\label{rip_par}
\end{table}

The univariate estimations for the periods gave similar results to Bitcoin's and Litecoin's. The shape parameter $\alpha$ is decreasing, while $\gamma$ is rising as time passes. The $\beta$ shows a significant skewness to the right, however the shift $\delta$ is always negative.

\begin{table}[H]
\centering
\begin{tabular}{|l|l|l|l|}
\hline
Ripple & \multicolumn{1}{c|}{1.} & \multicolumn{1}{c|}{2.} & \multicolumn{1}{c|}{3.} \\ \hline
Critical Value & 2.818 & 2.154 & 2.269 \\ \hline
Test statistic & 1.020 & 1.255 & 1.209 \\ \hline
\end{tabular}
\caption{The results of the Anderson-Darling tests for the distribution of Ripple logreturns.}
\label{AD_RIP}
\end{table}

The AD test didn't reject the null-hypothesis for any of the three periods, all the value of the test statistics are under the critical values chosen on 95\% confidence level. Unfortunately, many of the simulated test statistic values were non-interpretable again, but enough usable remained to evaluate the tests.

We performed the multivariate goodness of fit tests too with Kendall functions. In 3 dimensions, the empirical Kendall functions are calculated from the values 
\[
M_i=\frac{1}{n}\sum_{j\neq i}\bm{1}(X_j<X_i,Y_j<Y_i,Z_j<Z_i),\quad i=1\ldots n.
\] 
In words, for a given point triplet we have to count how many points fall under it within all three coordinates, then from the calculated values, we construct the empirical Kendall function of the distribution and we are ready for testing and for the simulation. 
The fitting of the distributions was calculated using three different point calibrations again, with 8, 16 and 32 points, now from the earlier mentioned circular cross section of the sphere (Figure \ref{circ}). In 3 dimensions this means we based the fit on ${3\choose 2}\cdot 8=24$, ${3\choose 2}\cdot 16=48$ and ${3\choose 2}\cdot 32=96$ points in total, which is a significant raise in the number of parameters. Since we have only selected three periods now, the total running time of the 

Cramér--von Mises  tests were way lower than before.

\begin{table}[H]
\centering
\begin{tabular}{c|c|c|c|c|c|c|}
\cline{2-7}
\multicolumn{1}{l|}{}        & \multicolumn{2}{c|}{${3\choose 2}\cdot 8$ points}                                                                                         & \multicolumn{2}{c|}{${3\choose 2}\cdot 16$ points}                                                                                        & \multicolumn{2}{c|}{${3\choose 2}\cdot 32$ points}                                                                                        \\ \hline
\multicolumn{1}{|c|}{Window} & \begin{tabular}[c]{@{}c@{}}Test \\ statistic\end{tabular} & \begin{tabular}[c]{@{}c@{}}Critical\\  value\end{tabular} & \begin{tabular}[c]{@{}c@{}}Test \\ statistic\end{tabular} & \begin{tabular}[c]{@{}c@{}}Critical \\ value\end{tabular} & \begin{tabular}[c]{@{}c@{}}Test\\  statistic\end{tabular} & \begin{tabular}[c]{@{}c@{}}Critical\\  value\end{tabular} \\ \hline
\multicolumn{1}{|c|}{1.}     & 2.142                                                     & 3.501                                                     & \cellcolor[HTML]{CB0000}4.802                             & 1.633                                                     & 1.453                                                     & 3.388                                                     \\ \hline
\multicolumn{1}{|c|}{2.}     & 1.052                                                     & 4.173                                                     & \cellcolor[HTML]{CB0000}2.077                             & 0.659                                                     & 0.460                                                     & 2.076                                                     \\ \hline
\multicolumn{1}{|c|}{3.}     & 0.996                                                     & 3.661                                                     & 1.278                                                     & 4.303                                                     & \cellcolor[HTML]{CB0000}0.890                             & 0.490                                                     \\ \hline
\end{tabular}
\caption{Results of the performed 
	Cramér--von Mises type test statistics with three different calibrations. Critical values were chosen based on 95\% significance level as before. The values with red background are the tests, where the null hypothesis had to be rejected.}
\label{kend3}
\end{table}
The results are interesting, however they are not really parallel to the results seen in Table \ref{kendallstat}. The estimations based on 16 points per circular cross sections of the sphere were the worst of all, two periods got absolutely rejected. The third period is not rejected, but it can be generally said that the values of the test statistic are all higher with 16 points than the other two approaches. Based on the results, estimation on 8 points was the best overall. The testing for the third period with 32 points is rejected too, but these results may have changed, if the simulation were done using more repetitions.

Since the spectral measure is now concentrated on the surface of a sphere, visualizations gets more difficult. Density plots are not feasible, so we show the spectral measures, but on simplified figures, showing every $\bm{\lambda}^{l,k}, l\neq k$, when the estimation was done on 32 points per circles.

\begin{figure}[H]
	\centering		
		\includegraphics[scale=0.35]{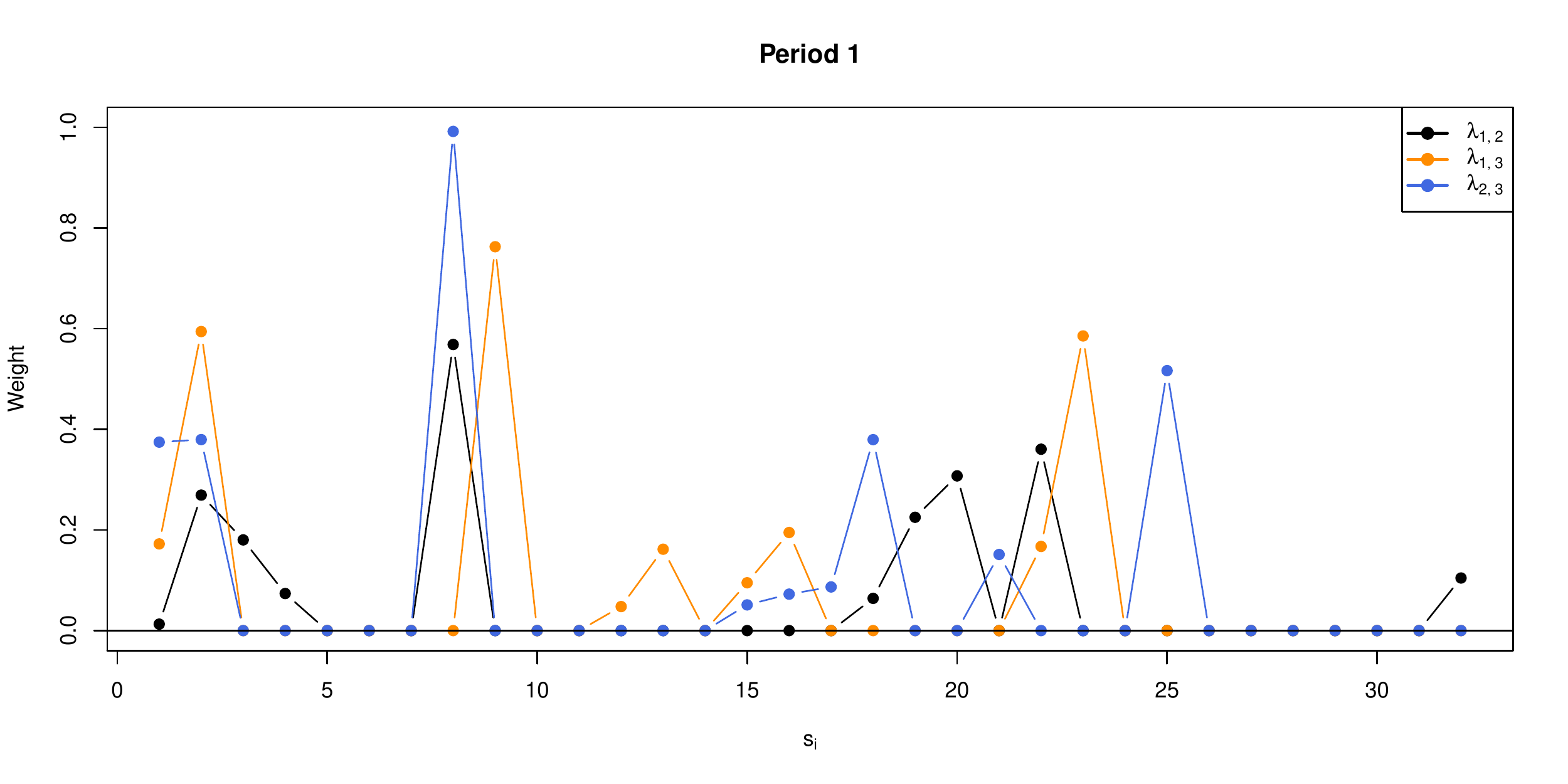}
\end{figure}
	\vspace{-14mm}
\begin{figure}[H]
	\centering		
		\includegraphics[scale=0.35]{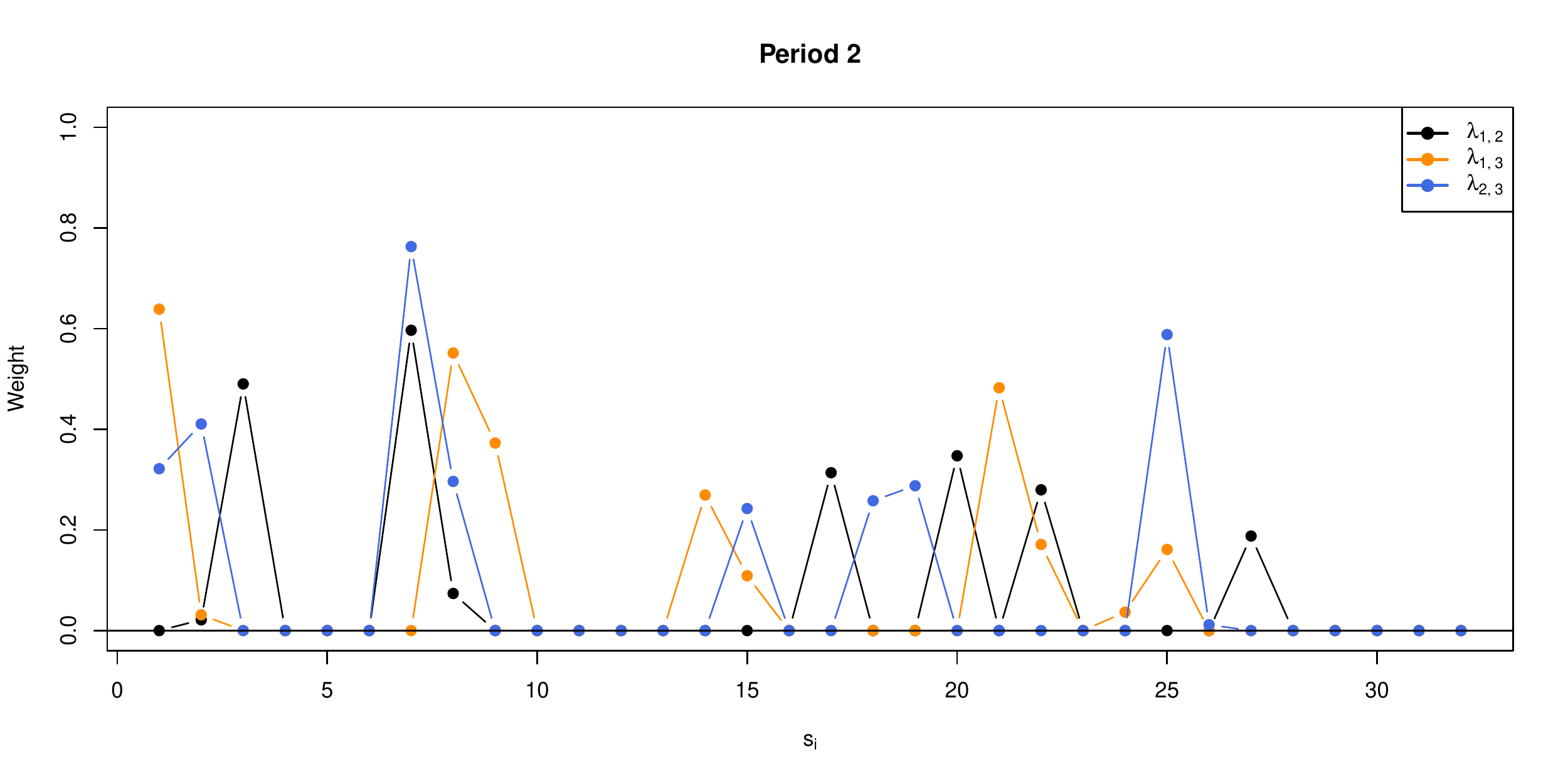}
\end{figure}
	\vspace{-14mm}
\begin{figure}[H]
	\centering		
		\includegraphics[scale=0.35]{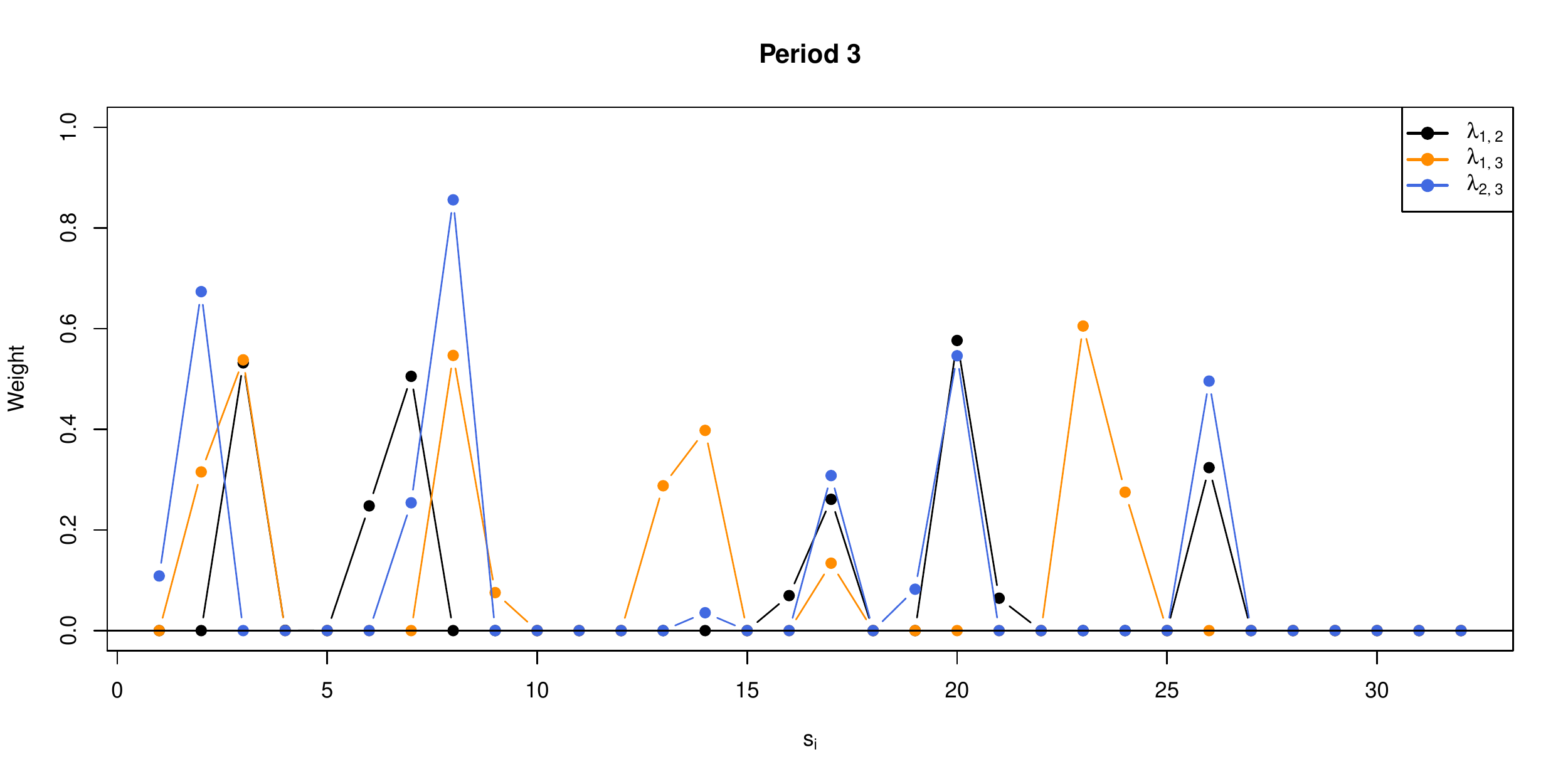}
		\caption{Estimated spectral measures (32 points per circles). Notation for the margins: 1--Bitcoin, 2--Litecoin, 3--Ripple.}
\end{figure}
It is visible, that every possible pair of cryptocurrency logreturns have overall similar dependence structures, so the three asset's price seems to react in a similar way to each other and to new information on the market. The change in the dependence structure is similar to what we could see on Figure \ref{dens32}: two dominant angles are present in the positive region of $\mathbb{R}^3$, caused by the weights $\bm{\lambda}_{1}^{l,k},\ldots,\bm{\lambda}_{10}^{l,k}$, which are getting closer to each other as time passes. The weights in the negative region of $\mathbb{R}^3$ are showing some realignment, focusing more onto a fewer density points, creating more dominant angles.

Despite the difficulties in visualizing the density, it is easy to calculate probabilities. Here, we take all three cryptocurrencies into consideration and estimate the probability
\[
P(\text{Ripple logreturn}<-10\% \mid \text{Bitcoin logreturn}<-10\%,\text{Litecoin logreturn}<-10\%)
\]
from the fitted normal and earlier fitted stable distributions. The results are a bit different though, as the probability calculated from the normal distributions are higher now, unlike in Table \ref{prob1}.
\begin{table}[H]
\centering
\begin{tabular}{|l|l|l|l|}
\hline
Period & \multicolumn{1}{c|}{1.} & \multicolumn{1}{c|}{2.} & \multicolumn{1}{c|}{3.} \\ \hline
Normal & 30.243\% & 22.018\% & 20.21\% \\ \hline
Stable & 87.903\% & 86.385\% & 63.481\% \\ \hline
\end{tabular}
\caption{Calculated conditional probabilities from the 3 dimensional normal and stable distributions, fitted to the logreturns.}
\label{prob2}
\end{table}
The probabilities based on the stable distributions are still significantly higher than the ones from the normal. It is interesting that for both, the probabilities decrease parallel to each other, but with different intensity.

\section{Comparison with other methods}

There are of course alternatives to using stable distributions for modelling cryptocurrency-returns. In the multivariate case copulas are very popular (see e.g. Bouyé, 2009 \cite{bo} for a review of their financial applications). It is recognised, that the famous parametric families usually do not capture well the dependencies between the coordinates. Extremal dependence is present for the $t$-copula, but its ellipticity is not observed in our case. 

Extreme value models are valid from a theoretical point of view, due to the Fisher-Tippet theorem, but in this case the maxima and the minima have to be modelled separately. It is also interesting to note that these estimations (e.g. the Hill estimator) give a value of $\alpha^*\sim 3$. $\alpha^*$ is the same parameter of the distribution as the shape $\alpha$ in the sense that both give the speed of decrease for $x\to\infty$: $F(x)~1/(x^{\alpha^*})$, so the tails do not look as heavy as based on the fitted stable model. 

\begin{figure}[H]
	\centering		
	\includegraphics[scale=0.35]{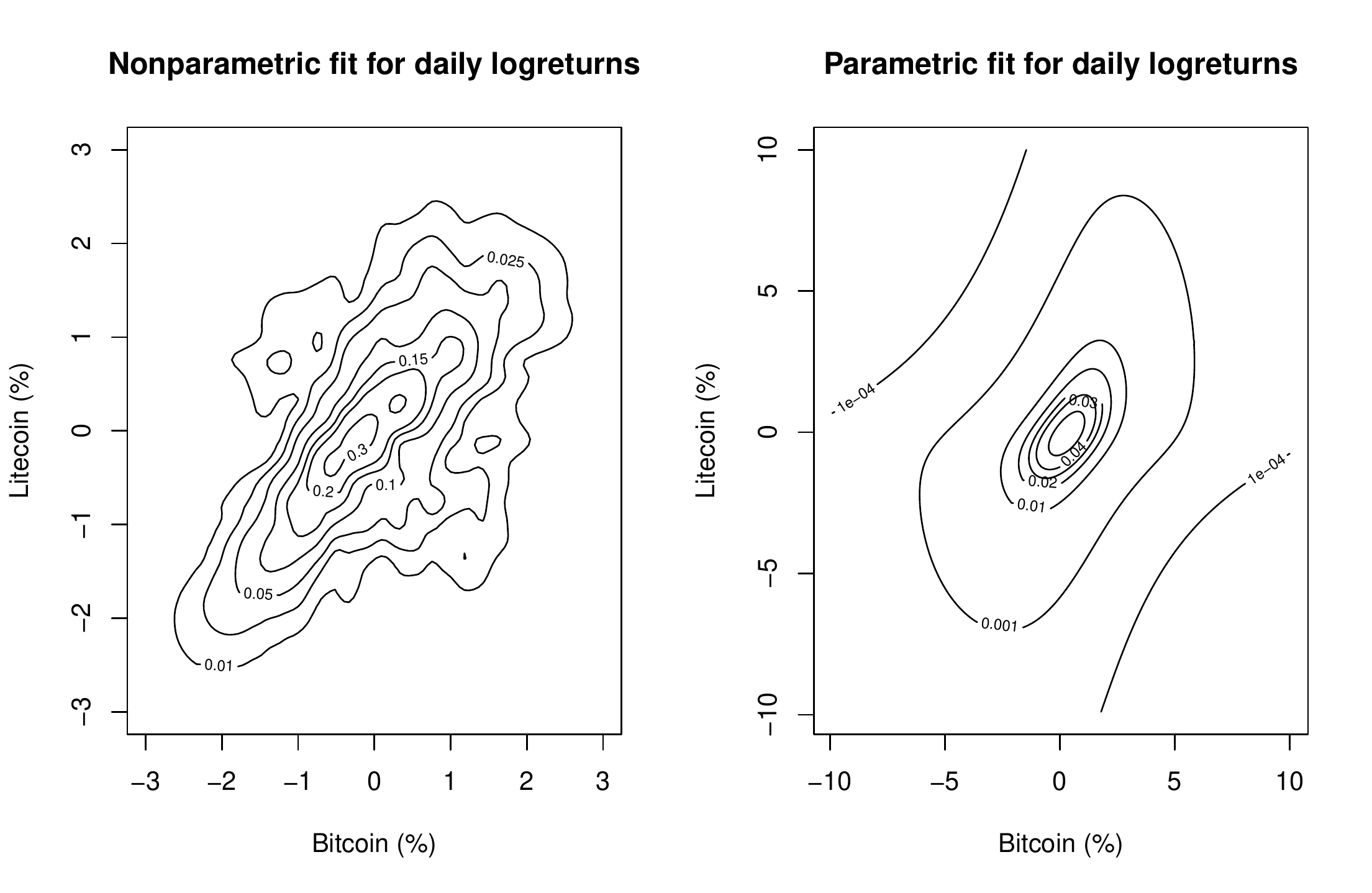}
	\caption{Comparison of nonparametric and parametric copula methods -- transformed to the original scale of the data}
	\label{cop}
\end{figure}

A very simple alternative might be a nonparametric copula estimator (based on the package {\tt{kdecopula}}) -- but it is no surprise that it cannot produce reasonable results, as here the procedure is based on the actually observed values. Figure \ref{cop} shows the results of the nonparametric copula estimation and the bivariate estimation calculated from stable marginals and $t_2$ copula, for the first period. It can be seen that the parametric modelling can catch the heavy tails, due to the marginal stable fit, but the coupling of heavy tailed margins with the not too strongly correlated dependence structure results in weak extremal dependence.

\newpage
\section{Summary}
This paper has shown the use of multivariate stable distributions in statistical modeling of heavy-tailed data by constructing a general multivariate estimation method. 

First let us bring attention to a few properties that we didn't mention before. Despite that we had promising results, there are a few problems with stable distributions in applications. First, when we are simulating from stable distributions with such low $\alpha$ that we could see before, there will be unusually big or small values in our samples. This is why it is necessary to use ML method for estimating $\alpha$, which usually doesn't underestimate $\alpha$. Second, we need sufficient amount of elements in the samples to be able to perform the univariate estimation as best we can. Low sample sizes may result in false parameters $\alpha$. Third, for the multivariate estimation, if the estimated $\alpha$ parameters of the marginal distributions differ too much, we shouldn't try to fit multivariate stable distribution. In this case, the pooled $\alpha^*$ would give absolutely false results, because every individual $\alpha$ should be close to each other. This is a consequence of the Proposition \ref{proj_tul}.

It turned out that the methods were applicable to daily logreturns of cryptocurrencies in 3 dimensions. We can say, that fitting stable distributions to logreturns of cryptocurrencies in the tested dimensions can be used very well, nevertheless the distribution family was rejected by a few authors, when modeled stock returns. Based on our results, stable distributions could be used for modeling the price changes of cryptocurrencies. We have also seen that for modelling dependencies in this very heavy-tailed case the multivariate stable distributions are much more realistic -- not only because of their theoretical advantage, but also on the much more realistic fits, compared to the more popular copulas.

\section*{Acknowledgment}
The project has been supported by the
European Union, co-financed by the
European Social Fund
(EFOP-3.6.3-VEKOP-16-2017-00002).

\newpage

\end{document}